\documentclass[]{spie}  

\usepackage[]{graphicx}

\graphicspath{{.}}

\usepackage{algorithm}
\usepackage{amstext}
\usepackage{amsmath} 
\usepackage{amssymb}

\title{ k-core decomposition: a tool for the visualization of large scale networks}

\author{ Ignacio Alvarez-Hamelin\supit{a}, Luca Dall'Asta\supit{a},
Alain Barrat\supit{a} and Alessandro Vespignani\supit{b}
\skiplinehalf
\supit{a}LPT (UMR du CNRS 8627),
Universit\'e de Paris-Sud,
         91405 ORSAY Cedex France; \\
\supit{b}School of Informatics, Indiana University, Bloomington, IN 47408, USA.
}

\authorinfo{Further author information: 
I.A-H. is also with Facultad de Ingenier\'{\i}a,
         Universidad de Buenos Aires, Paseo Col\'on 850,
         C 1063 ACV Buenos Aires, Argentina.
\\I.A.-H.: E-mail: Ignacio.Alvarez-Hamelin@lri.fr, 
Telephone: +33-1-6915-8222
}

\begin{document}

\maketitle

\begin{abstract} 
  We use the $k$-core decomposition to visualize large scale complex
  networks in two dimensions. This decomposition, based on a 
  recursive pruning of the least
  connected vertices, allows to disentangle the hierarchical structure of
  networks by progressively focusing on their central cores. By using
  this strategy we develop
  a general visualization algorithm that can be used to compare the 
  structural properties of various networks and
  highlight their hierarchical structure. 
  The low computational complexity of the algorithm, ${\cal O}(n+e)$, where $n$ 
  is the size of the network, and $e$ is the number of edges,
  makes it suitable for the visualization
  of very large sparse
  networks.  We apply the proposed visualization tool to several real
  and synthetic graphs, showing its utility in finding specific
  structural fingerprints of computer generated and real
  world networks.
\end{abstract} 
 
\section{Introduction}\label{intro} 

In recent times, the possibility of accessing, handling and mining
large-scale networks datasets has revamped the interest
in their investigation and theoretical characterization along with
the definition of new modeling frameworks. In particular, mapping
projects of the World Wide Web (WWW) and the physical Internet 
offered the first chance
to study topology and traffic of large-scale networks. Gradually
other studies followed describing population networks of practical
interest in social science, critical infrastructures and
epidemiology~\cite{Barabasi:2000,Amaral:2000,mdbook,psvbook}.  
The study of large scale networks, however, faces us with an array of 
new challenges. The definitions of centrality, hierarchies and
structural organizations are hindered by the large size of these
networks and the complex interplay
of connectivity patterns, traffic flows and geographical, social and
economical attributes characterizing their basic elements. 
In this context, a large research effort is devoted to provide effective
visualization and analysis tools able to cope with graphs whose size
may easily reach millions of vertices.
 
In this paper, we propose a
visualization algorithm based on the $k$-core decomposition 
able to uncover in a two-dimensional layout several 
topological and hierarchical properties of large scale networks.  
The $k$-core decomposition~\cite{Batagelj02} consists in identifying 
particular subsets of the graph, called $k$-cores, each one 
obtained by recursively removing all the vertices of degree 
smaller than $k$, until the degree of all remaining vertices 
is larger than or equal to $k$. Larger values of ``coreness'' clearly 
correspond to vertices with larger degree and more central position 
in the network's structure.

When applied to the
graphical analysis of real and computer-generated networks, this
visualization tool allows the identification of networks'
fingerprints, according to properties such as  hierarchical
arrangement, degree correlations and centrality.  The distinction
between networks with seemingly similar properties is achieved
by inspecting the different layouts generated by the visualization 
algorithm. In addition, the running time of the algorithm grows 
only linearly with the size of the network, granting the scalability
needed for the visualization of very large networks. 
The proposed visualization algorithm appears therefore as a
convenient method for the general analysis of large scale complex
networks and the study of their architecture. The presented
visualization algorithm is publicly available~\cite{LANET-VI}.
 
The paper is organized as follows: after a brief survey on 
$k$-core studies (section~\ref{related}), the basic
definitions are introduced in section~\ref{sec:k-core}; the graphical
algorithms are exposed in section~\ref{graph_rep} along with the basic
features of the visualization layout. Section~\ref{fingprint} shows
how the visualizations obtained with the present algorithm may be used
for network fingerprinting, while
section~\ref{results} is devoted to the application of the algorithm
to the visualization of various real and computer-generated networks.
 
\section{Related work}\label{related} 
%
While a large number of algorithms aimed at the visualization of large scale
networks have been developed (e.g., see~\cite{COSIN}), only a 
few consider explicitly the $k$-core decomposition.
Vladimir Batagelj {\em et al.}~\cite{Batagelj99}
studied the $k$-cores decomposition applied to visualization problems,
introducing some graphical tools to analyse the cores, mainly based on
the visualization of the adjacency matrix of certain $k$-cores.
To the best of our knowledge, the algorithm presented by 
Baur {\em et al.} in the paper
``Drawing the AS Graph in 2.5 Dimensions''~\cite{Baur04}, is the only one
completely based on a $k$-core analysis and directly targeted at the
study of large information networks. 
This  algorithm uses a 
spectral layout to place vertices having the largest coreness.  A
combination of barycentric and iteratively directed-forces allows
to place the vertices of each $k$-shell, in decreasing order. Finally,
the network is drawn in three dimensions, using the $z$ axis to place
each coreness set in a distinct horizontal layer. It is important to
stress that the spectral layout is not able to distinguish two or more
disconnected components.  
The algorithm by Baur {\em et al.} is also 
tuned for representing AS graphs and its
total complexity depends on the size of the highest $k$-core
(see~\cite{brandes03} for more details on spectral layout), making the
computation time of this proposal largely variable.
In this respect, the algorithm presented here is considerably 
different in that it can represent networks in which  $k$-cores
are composed by several connected components.
Another difference is that representations in 2D are more suited for 
information visualization than other 
representations (see~\cite{2Dvs3D} and references therein). 
Finally, the algorithm parameters can be universally 
defined (see section~\ref{results}), 
yielding a fast and general tool for analyzing all types of networks.

It is interesting to note that the notion of $k$-cores 
has been recently used in biologically related contexts, 
where it was applied to the analysis of protein interaction
networks \cite{bader03} or in the prediction of protein functions
\cite{gnome-info03,Wuchty05}. A further interesting application in the area
of networking has been provided by Gaertler {\em et
al.}~\cite{Gaertler04}, where the $k$-core decomposition is used 
for filtering out peripheral Autonomous Systems (ASes) in the case of
Internet maps. 

\section{$k$-core decomposition: main definitions}\label{sec:k-core} 
 
Let us consider a graph $G=(V,E)$ of $|V|=n$ vertices and $|E|=e$ edges;
a $k$-core is defined as follows~\cite{Batagelj02}:
 
\begin{definition}\label{k-core} 
A subgraph $H = (C,E|C)$ induced by the set $C\subseteq V$ is a {\em
$k$-core} or a core of order $k$ iff $\forall v \in C: {\tt
degree}_H(v)\geq k$, and H is the maximum subgraph with this property.
\end{definition} 
 
A $k$-core of $G$ can therefore be obtained by recursively removing all the 
vertices of degree less than $k$, until  all 
vertices in the remaining graph have at least degree $k$. 

Furthermore, we will use the following definitions: 
 
\begin{definition}\label{coreness} 
A vertex $i$ has {\em coreness} $c$ if it belongs to the 
$c$-core but not to $(c+1)$-core. We denote by $c_i$ the coreness of 
vertex $i$. 
\end{definition} 
 
\begin{definition}\label{set_coreness} 
A shell $C_c$  
is composed by all the vertices whose coreness is $c$. The maximum 
value $c$ such that $C_c$ is not empty is denoted $c_{\max}$. The $k$-core 
is thus the union of all shells $C_c$ with $c \ge k$. 
\end{definition} 
 
\begin{definition}\label{cluster} 
Each connected set of vertices having the same coreness $c$ 
is a {\em cluster} $Q^c$.  
\end{definition} 
Each shell $C_c$ is thus composed by clusters $Q_m^c$, such that 
$C_c=\cup_{1\leq m \leq q_{\max}^c} Q^c_m$, where $q_{\max}^c$ is 
the number of clusters in $C_c$.  

\begin{figure}[t]
\begin{center}
\begin{tabular}{|c|}
\hline
\includegraphics[width=30mm,angle=-90]{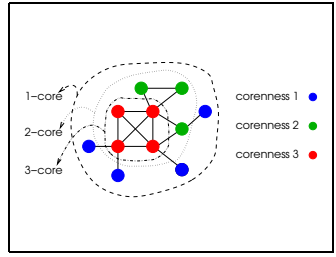}\\
\hline
\end{tabular}
\end{center}
\caption{Sketch of the $k$-core decomposition for a small
graph. Each closed line contains the set of vertices belonging to a
given $k$-core, while colors on the vertices distinguish different
$k$-shells.}
\label{k-core1}
\end{figure}

In Fig.\ref{k-core1} we report a simple illustration of k-core
decomposition of a connected graph and its visual rendering. Every
vertex of a connected graph belongs to the $1$-core. In
Fig.\ref{k-core1}, we have highlighted the different cores using closed
lines of different types. A dashed line encloses all the vertices in
the $1$-core (the entire graph).Then, all vertices of degree $d < 2$
are recursively cut out. In Fig.\ref{k-core1} all these vertices are
colored in blue. The other vertices maintain a degree $d \geq 2$ also
after the pruning of the blue ones, therefore they are not
eliminated. The remaining vertices form the $2$-core, enclosed by a
dotted line. Further pruning allows to identify the innermost set of
vertices, the $3$-core. One can check that all red vertices in
Fig.\ref{k-core1} have internal degree (i.e. between red vertices) at
least $3$. This core is highlighted by a dash-dotted line. This simple
process and its visual rationalization is at the basis of the
construction of our visualizations algorithm and layout.
   
\section{Graphical representation}\label{graph_rep} 
 
The visualization algorithm we propose places vertices in $2$
dimensions, the position of each vertex
depending on its coreness and on the coreness of its neighbors. 
A color code allows for the identification of core
numbers, while the vertex's original degree is provided by
its size that depends logarithmically on the degree. For the sake of
clarity, our algorithm represents a small percentage of
the edges, chosen uniformly at random.
As mentioned, a central role in our visualization method is played by
multi-components representation of $k$-cores. In the most general
situation, indeed, the recursive removal of vertices having degree less
than a given $k$ can break the original network into various connected
components, each of which might even be once again broken by the
subsequent decomposition.  Our method takes into account this
possibility, however we will first present the
algorithm in the simplified case (Table~\ref{A:algorepre}), in which
none of the $k$-cores is fragmented.  Then, this algorithm will be
used as a subroutine for treating the general case
(Table~\ref{A:algorepre2}).
 
\subsection{Drawing algorithm for $k$-cores with single connected component} 
 
The network under study is represented by a graph $ G=\{V,E\}$, where
$V$ is the set of vertices and $E$ is the set of links. 

\noindent{\bf $k$-core decomposition}.
The coreness of
each vertex is computed (according to the procedure described in
section~\ref{sec:k-core}) and stored in a vector ${\cal C}$, along
with the shells $C_c$ and the maximum coreness value $c_{max}$.
Each shell is then decomposed into clusters $Q^c_m$ of
connected vertices, and each vertex $i$ is labeled by its coreness $c_i$
and by a number $q_i$ representing the cluster it belongs to.

\noindent{\bf The two dimensional graphical layout}. The visualization 
is obtained assigning to each vertex $i$ a
couple of polar coordinates ($\rho_i,\alpha_i$): the radius
$\rho_i$ is a function of the coreness of the vertex $i$ and
of its neighbors; the angle $\alpha_i$ depends on the
cluster number $q_{i}$.  In this way, $k$-shells are displayed as
layers with the form of circular shells, the innermost one
corresponding to the set of vertices with highest coreness. A vertex $i$
belongs to the $c_{\max}-c_i$ layer from the center.
 
More precisely, $\rho_i$ is computed according to the following 
formula: 
\begin{equation} \label{radius} 
\rho_i = (1 - \epsilon ) ( c_{\max}-c_i ) +  
\frac{\epsilon}{|V_{c_j\geq c_i}(i)|}  
\quad \sum_{j \in V_{c_j\geq c_i}(i)} ( c_{\max}-c_j)  
\enspace,
\end{equation} 
$V_{c_j\geq c_i}(i)$ is the set of neighbors of $i$ having coreness 
$c_j$ larger or equal to $c_i$. The parameter $\epsilon$ controls the 
possibility of rings overlapping, and is one of the only three external 
parameters required to tune image's rendering. 
 
Inside a given shell, the angle $\alpha_i$ of a vertex $i$ 
is computed as follow: 
\begin{equation} 
\alpha_i = 2 \pi \sum_{1\leq m < q_i} \frac{|Q_m|}{|C_{c_i}|} +  
\mbox{\Large \bf N} \left( \frac{|Q_{q_i}|}{2 |C_{c_i}|} \;\; ,  
\;\; \pi \cdot \frac{|Q_{q_i}|}{|C_{c_i}|} \right)  
\enspace,\label{angle} 
\end{equation} 
where $Q_{q_i}$ and $C_{c_i}$ are respectively the cluster $q_{i}$ 
and $c_{i}$-shell the vertex belongs to, {\large \bf N} is a
normal distribution of mean $\frac{|Q_{q_i}|}{2 |C_{c_i}|}$ and width
$2 \pi \cdot \frac{|Q_{q_i}|}{|C_{c_i}|}$.  Since we are interested in
distinguishing different clusters in the same shell, the first term on
the right side of Eq.~\ref{angle}, referring to clusters with $m<q_i$,
allows to allocate a correct partition of the angular sector to each
cluster.  The second term on the right side of Eq.~\ref{angle},
on the other hand, specifies a random position for the vertex $i$ in the sector
assigned to the cluster $Q_{q_i}$.
 
\noindent{\bf Colors and size of vertices}. Colors are assigned 
according to the coreness: vertices with coreness $1$ are violet, 
and the maximum coreness vertices are red, 
following the rainbow color scale.  Finally, the
diameter of each vertex corresponds to the logarithm of its degree,
giving a further information on vertex's properties.  Note that the
vertices with largest coreness are placed uniformly in a disk of radius
$u$, which is the unit length ($u$ equals $1$ for this reduced
algorithm).
 
The complete algorithm is presented in Table~\ref{algorepre}. In
particular, vector ${\cal Q}$ collects the cluster numbers $\{q_i\}$
of all vertices, and table ${\cal T}$ contains the following pair of
elements, indexed by the coreness $c$ and cluster label $q$
\begin{equation}
{\cal T}(c,q)=\left(\sum_{1\leq m < q} \frac{|Q_m|}{|C_{c}|}\;,\;\frac{|Q_{q}|}{|C_{c}|}\right) 
\enspace. \label{matrixT}
\end{equation}
These input quantities, used in Eq.~\ref{angle}, can be computed during the
$k$-core decomposition, when the cluster labels are assigned.
\begin{table} 
\begin{algorithm}{}
\label{A:algorepre} 
\\ 
input: vectors of coreness ${\cal C}$ and cluster ${\cal Q}$, 
and ${\cal T}$, indexed by vertex $i$ \\ 
for each vertex $i$ do \> \\ 
    if $c_i==c_{\max}$ then \> \\ 
        set $\rho_i$ and $\alpha_i$  according to a uniform  
distribution in the disk of radius $u$ 
($u$ is the core representation unit size)  
        \< \\ 
    else \> \\ 
                set $\rho_i$ and $\alpha_i$ according to Eqs.~\ref{radius} and~\ref{angle} 
        \<  
    \<\\ 
return $\rho$ and $\alpha$ vectors 
\end{algorithm} 
\caption{Algorithm for representing networks using $k$-cores decomposition} 
\label{algorepre} 
\end{table}

\subsection{Extended algorithm using $k$-cores components} 
 
The algorithm presented in the previous section can be used as the
basic routine to define an  extended algorithm aimed at the
visualization of networks for which some $k$-cores are fragmented;
i.e. made by more than one connected component.  
This issue is solved by assigning to each connected component of a
$k$-core a center and a size, which depends on the relative sizes of
the various components. Larger components are put closer to the global
center of the representation (which has Cartesian coordinates
$(0,0)$), and have larger sizes.
 
The algorithm begins with the center at the origin $(0,0)$. Whenever a
connected component of a $k$-core, whose center $p$ had coordinates
$(X_p,Y_p)$, is broken into several components by removing all vertices
of degree $k$, i.e. by applying the next decomposition step, a new
center is computed for each new component. The center of the component
$h$ has coordinates $(X_h , Y_h)$, defined by
\begin{equation} 
X_h  =  X_p + \delta (c_{\max} - c_h) \cdot u_p \cdot \varrho_h 
\cdot \cos ( \phi_h ) \label{x_center} \ ; \ \ 
 Y_h  =  \; Y_p + \delta 
(c_{\max} - c_h) \cdot u_p \cdot \varrho_h \cdot \sin ( \phi_h )\enspace, 
\end{equation}
where $\delta$ scales the distance between components, $c_{\max}$ is
the maximum coreness and $c_h$ is the core number of component $h$
(the components are numbered by $h=1,\cdots,h_{max}$ in an 
arbitrary order), $u_p$
is the unit length of its parent component, $\varrho_h$ and $\phi_h$
are the radial and angular coordinates of the new center with respect
to the parent center $(X_p,Y_p)$. We define $\varrho_h$ and $\phi_h$
as follows:
\begin{equation} 
 \varrho_h  =  1 - \frac{|S_h|}{\sum_{1\leq j \leq h_{max}} |S_j|} \ ;\ \ 
 \phi_h  =  \phi_{ini} + \frac{2 \pi} 
{\sum_{1\leq j \leq h_{max}} |S_j|} 
\sum_{1\leq j \leq h} |S_j| \label{cmp}\enspace, 
\end{equation} 
where $S_h$ is the set of vertices in the component $h$, $\sum_j
|S_j|$ is the sum of the sizes of all components having the same
parent component. In this way, larger components will be closer to the
original parent component's center $p$.
 
The angle $\phi_h$ has two contributions.  The initial angle 
$\phi_{ini}$ is chosen uniformly at random\footnote{Note that if the 
$\phi_{ini}$ is fixed, all the centers of the various 
components are aligned in the final representation.}, 
while the angle sector is the sum of component angles whose number is 
less than or equal to the actual component number $h$.

Finally, the unit length $u_h$ of a component $h$ is computed as 
\begin{equation} 
  u_h = \frac{|S_h|}{\sum_{1\leq j \leq h_{max}} |S_j|} 
\cdot u_p 
  \enspace, \label{u_h} 
\end{equation} 
where $u_p$ is the unit length of its parent component.  
Larger unit length and size are therefore attributed to  
larger components. 
 
For each vertex $i$, radial and angular coordinates are computed 
by equations~\ref{radius} and~\ref{angle} as in the previous algorithm. 
These coordinates are then considered as relative to the center $(X_h,Y_h)$ of 
the component to which $i$ belongs. The position of $i$ is thus given by 
\begin{equation} 
x_i  =  X_h + \gamma \cdot u_h \cdot \rho_i \cdot \cos ( \alpha_i ) 
; \ \ 
y_i =  \; Y_h + \gamma \cdot u_h \cdot \rho_i 
\cdot \sin ( \alpha_i ) \label{y_vertex_II} 
\end{equation} 
where $\gamma$ is a parameter controlling the component's diameter. 
 
The global algorithm is formally presented in
Table~\ref{A:algorepre2}.
The main loop is composed by the following functions.   
First, the function \mbox{\{$(end,{\cal C})\leftarrow${\tt make\_core
$k$}\}} recursively removes all vertices of degree $k-1$, obtaining the
$k$-core, and stores into ${\cal C}$ the coreness $k-1$ of the removed
vertices.  The boolean variable $end$ is set to $true$ if the $k$-core is
empty, otherwise it is set to $false$.  The function \mbox{\{$({\cal
Q},{\cal T})\leftarrow$ {\tt compute\_clusters $k-1$}\} } operates the
decomposition of the $(k-1)$-shell into clusters, 
storing for each vertex
the cluster label into the vector ${\cal Q}$, and filling table
${\cal T}$ (see Eq.~\ref{matrixT}).  The possible decomposition of the
$k$-core into connected components is determined by function
\mbox{\{${\cal S}\leftarrow$ {\tt compute\_components $k$}\}}, that
also collects into a vector ${\cal S}$ the number of vertices contained
in each component. At the following step, functions
\mbox{\{$(X,Y)\leftarrow${\tt compute\_origin\_coordinates\_cmp $k$}\}}
and
\mbox{\{$U\leftarrow${\tt compute\_unit\_size\_cmp $k$}\}} get,
respectively, the center and size of each component of the $k$-core,
gathering them in vectors $X$, $Y$ and $U$.  Finally, the coordinates
of each vertex are computed and stored in the vectors ${\cal X}$ and
${\cal Y}$.
\begin{table} 
\begin{algorithm}{}\label{A:algorepre2} 
\\  
$k:=1$ and $end:={\tt false}$ \\  
while {\tt not} $end$ do \>\\  
  $(end,{\cal C})\=${\tt make\_core $k$}\\  
  if $k > 1$ $({\cal Q},{\cal T})\=${\tt compute\_clusters $k-1$} \\  
  ${\cal S}\=$ {\tt compute\_components $k$} \\  
  $(X,Y)\=${\tt compute\_origin\_coordinates\_cmp} $k$  
(Eqs.~\ref{x_center} and~\ref{cmp}) \\  
  $U\=${\tt compute\_unit\_size\_cmp} $k$ (Eq.~\ref{u_h}) \\  
  $k:=k+1$ \< \\ 
$(\rho,\alpha)\=${\bf Algorithm~\ref{A:algorepre}} with  
${\cal C}$, ${\cal Q}$, ${\cal T}$ and $U$\\  
$({\cal X},{\cal Y})\=${\tt compute\_final\_coordinates}  
$\rho\; \alpha\; U \; X\; Y$ (Eq.~\ref{y_vertex_II}) 
\end{algorithm} 
\caption{Extended algorithm for the representation of networks using $k$-cores 
decomposition} 
\label{algorepre2} 
\end{table} 
 
\noindent{\bf Algorithm complexity.} 
The $k$-core decomposition can be computed using the algorithm of 
Batagelj and Zaversnik~\cite{Batagelj03}. 
Two steps are necessary to perform the $k$-core decomposition of a
graph. First a list of the vertices with their respective neighbors is
prepared. The recursive pruning algorithm is applied. 
Building the list of
$n$ vertices with their degree takes a time ${\cal O}(n)$. Starting from
the lowest degree value $k_{min}$, all the vertices of degree equal to
$k_{min}$ are then recursively cut out. Pruning a neighbor $j$ of a
vertex $i$ of degree $k_{min}+1$ means that the degree of $i$ is
decreased to $k_{min}$, so that $i$ is subsequently pruned as well.
This is what is meant by the expression ``recursively cutting
out''. The first $k$-shell (of coreness $k_{min}$) contains all vertices
removed during this process.
When the remaining graph does not contain any vertex of
degree $k_{min}$, the algorithm repeats the procedure by recursively
removing vertices of degree $k_{min}+1$, thus constructing the
$k$-shell of coreness $k_{min}+1$.  
This process is repeated until no vertices are left, 
obtaining in this way the successive $k$-shells.
The construction of the $k$-shells takes a time 
time ${\cal O}(e)$ (where $e$ is the number of edges), 
because removing a vertex implies cutting the edges between this
vertex and its neighbors. The
building of all coreness sets thus implies that all edges are removed
one after the other in the process. In summary, the total time to
perform the decomposition is ${\cal O}(n+e)$.  In order to build the
clusters, each vertex should verify the coreness of its neighbors, 
which takes $2\cdot e$ steps in the worst case.
Finally, the total time complexity is ${\cal O}(n+e)$ for a general
graph. This makes the algorithm very efficient for sparse graphs
where $e$ is of order $n$.
 
\begin{figure}[t]
\begin{center}
\begin{tabular}{|c|} 
\hline
\includegraphics[angle=-90,width=75mm]{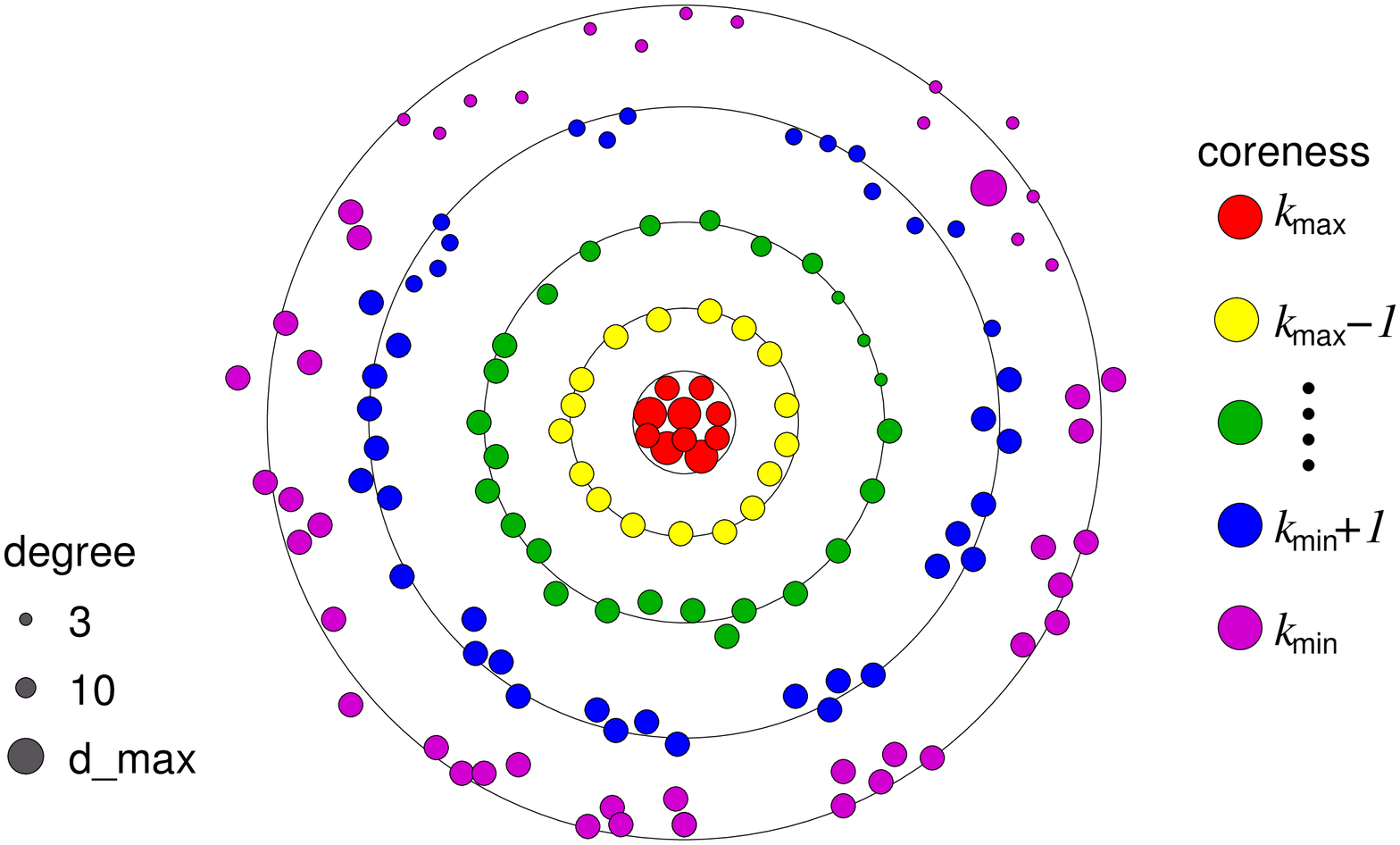}
\includegraphics[angle=-90,width=75mm]{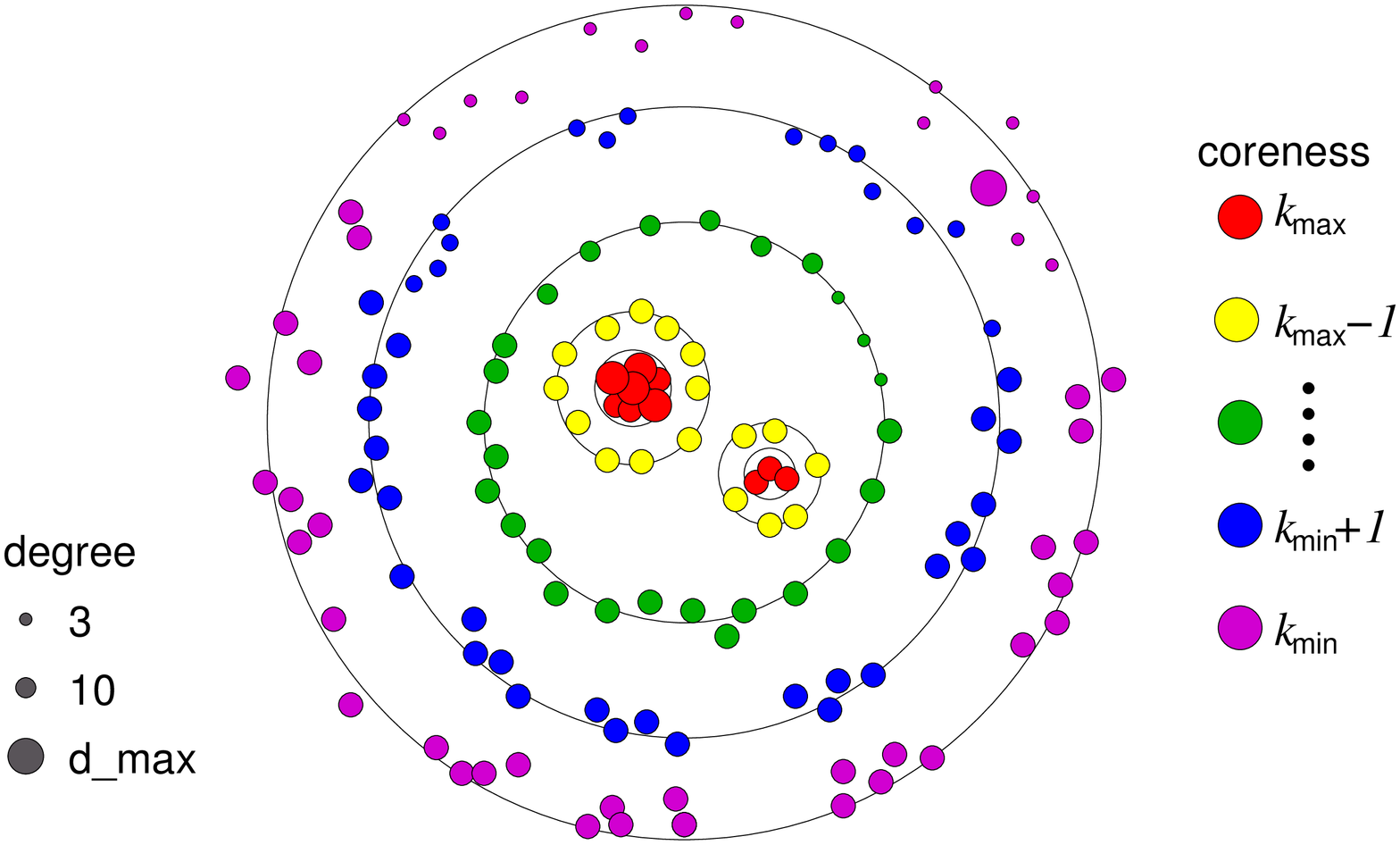}\\
\hline
\end{tabular}
\end{center}
\caption{
The two drawings show the 
structure of a typical layout in two important cases: on the
left, all $k$-cores are connected; on the right, some $k$-cores are composed
by more than one connected component.
The vertices are arranged in a series of concentric shells, each one 
corresponding to a particular $k$-shell. The diameter of the shell depends
on both the coreness value and, in case of multiple components (right) also
on the relative fraction of vertices belonging to the different components.
The color of the vertices corresponds to their coreness value, while their
size is logarithmically proportional to their original degree, as shown by the
scale going from the minimal to the maximal degree.
} 
\label{k-core2}
\end{figure}

\subsection{Basic features  of the visualization's layout}

The main features of the layout's structure obtained with the above
algorithms are visible in
Fig.\ref{k-core2} where, for the sake of simplicity, we do not show any edge.
The leftmost panel displays the case in which all $k$-cores have a single
component, while in the rightmost one an example of $k$-core fragmentation is
reported. Indeed, it is possible that, during the pruning procedure, the
remaining nodes forming a $k$-core do not belong to the same connected
component. When such a fragmentation occurs, the algorithm computes the multiple
components of the core and displays all of them in a coherent way.

$\bullet$
The {\bf visualization's layout} is {\em two-dimensional},
  composed of a series of concentric {\em circular} {\em shells}
  (see the five different shells in Fig.\ref{k-core2}).

$\bullet$ Each {\bf shell} corresponds to a single {\em coreness} value and
  all vertices in it are therefore drawn with the {\em same color}.
  A {\bf color scale} allows to distinguish different {\em coreness}
  values: in the layouts, as in Fig.\ref{k-core2}, the violet is used
  for the minimum value of coreness $k_{min}$, then nuances of blue, green and
  yellow compose a graduated scale for higher and higher coreness values up to
  the maximum value $k_{max}$ that is colored in red.

$\bullet$ The {\bf diameter} of each $k$-shell depends on the
  {\em coreness} value $k$, and is proportional to $k_{max}-k$
  (In Fig.\ref{k-core2}, the position of each shell
  is identified by a circle having the corresponding diameter). The presence
  of a trivial order relation in the coreness values ensures that all shells
  are placed in a concentric arrangement. On the other hand, when a $k$-core
  is fragmented in two or more components, the diameters of the different
  components depend also on the relative number of vertices belonging to each
  of them, i.e. the fraction between the number of vertices belonging to that
  component and the total number of vertices in that coreness set.  This is a
  very important information, providing a way to distinguish between multiple
  components at a given coreness value. Looking at the two central components
  for high coreness values in Fig.\ref{k-core2} (right), we immediately
  realize that the bigger one contains a larger fraction of vertices.

$\bullet$
Finally, the {\bf size} of each node is proportional to the
  {\em original degree} of that vertex; we use a logarithmic scale for
  the size of the drawn bullets.

\section{Network fingerprinting}
\label{fingprint}

The $k$-core decomposition peels the network layer by layer, revealing the
structure of the different shells from the outmost one to the more internal
ones. The algorithm provides a direct way to distinguish the network's
different hierarchies and structural organization
by means of some simple quantities: the radial width of the shells, the
presence and size of clusters of vertices in the shells, the correlations
between degree and coreness, the distribution of the edges interconnecting
vertices of different shells, etc.  The following features are useful
to extract this structural information out of the visualization. We
also highlight the role of the parameters $\epsilon$, $\delta$ and
$\gamma$ of the visualization algorithms in helping to determine the
structural characteristics of the visualized network.
\begin{figure}[t]
\begin{center}
\begin{tabular}{|c|}
\hline
\includegraphics[angle=-90,width=75mm]{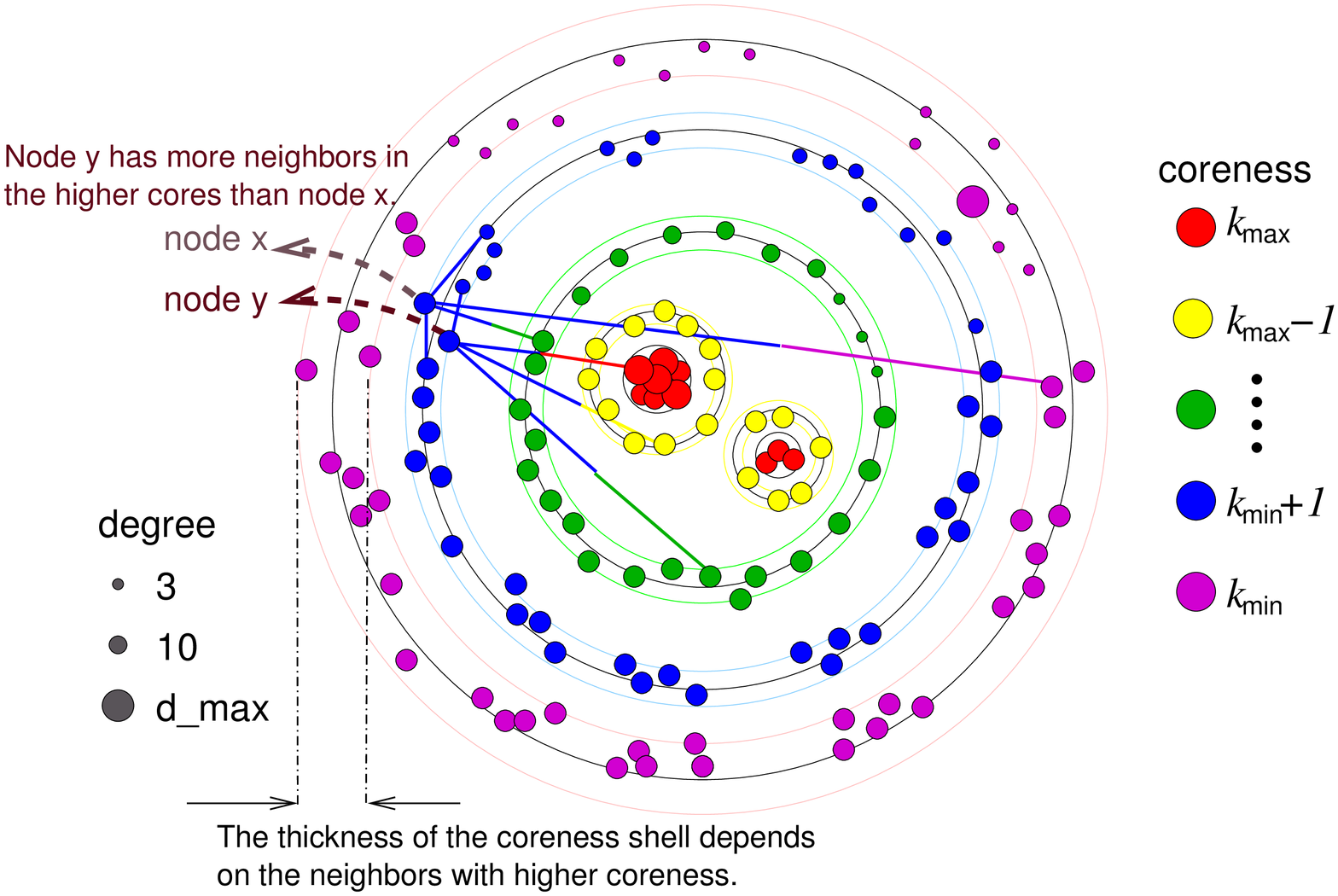} 
\includegraphics[angle=-90,width=75mm]{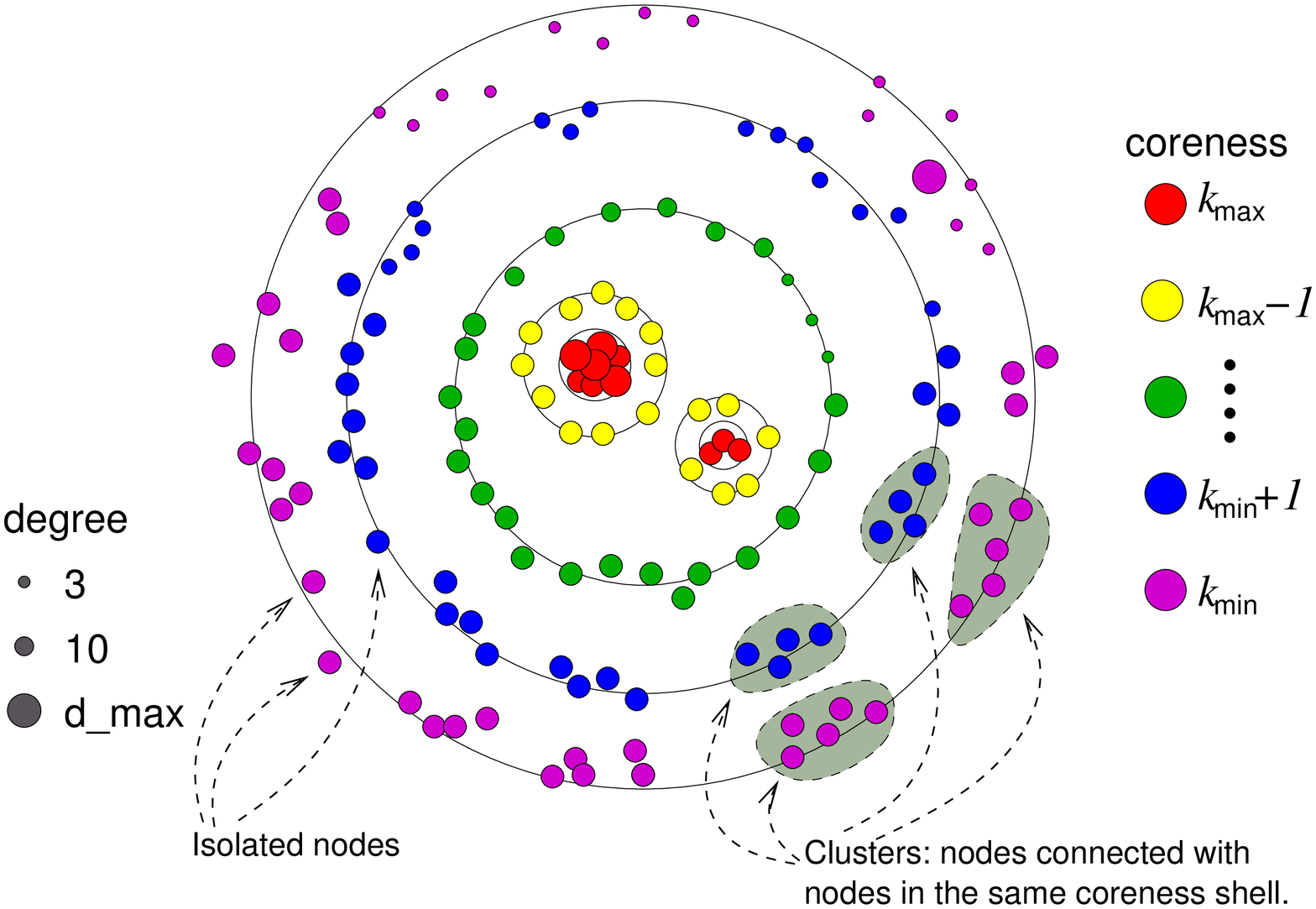} \\
\hline
\end{tabular}
\end{center}
\caption{ 
The figure on the left shows that each shell has a certain radial 
width. This width depends on the correlation's 
properties of the vertices in the shell. The dashed lines in the figure 
point out the width of the outmost shell, that corresponds to the lowest 
$k$-shell. In the second shell, we have pinpointed two nodes $x$ and $y$. 
The node $y$ is more internal than $x$ because a larger part of its neighbors 
belongs to higher $k$-shells compared to $x$'s neighbors. Indeed, $y$ has 
three links to nodes of higher coreness, while $x$ has only one.
The figure on the right shows the clustering 
properties of nodes in the same $k$-shell. In each $k$-shell, 
nodes that are directly connected between them (in the original graph) 
are drawn close one to the other, as in a cluster. Some of these sets 
of nodes are circled and highlighted in gray. Three examples of isolated 
nodes are also indicated; these nodes have no connections with the 
others of the same shell.
}
\label{k-core3}\label{k-core4}
\end{figure}

{\bf 1)}
\textit{\textbf{Shells Width:}} In the graph
  representations the width can change considerably from shell to shell.  The
  thickness of a shell depends on the coreness properties of the neighbors of
  the vertices in the corresponding coreness set. For a given shell-diameter
  (corresponding to the black circle in the median position of shells in
  Fig.\ref{k-core3}), each vertex can be placed more internal or more external
  with respect to this reference line. Nodes with more neighbors in higher
  coreness sets are closer to the center and viceversa, as shown in
  Fig.\ref{k-core3}. Node $y$ is more internal than node $x$ because it has
  three edges towards higher coreness nodes compared to the single edge
  emerging from $x$ towards inner shells.
  The maximum thickness of the shells is controlled by the $\epsilon$ 
  parameter (Eq.~\ref{radius}). 
\begin{figure}[t]
\begin{center}
\begin{tabular}{|c|}
\hline
\includegraphics[angle=-90,width=75mm]{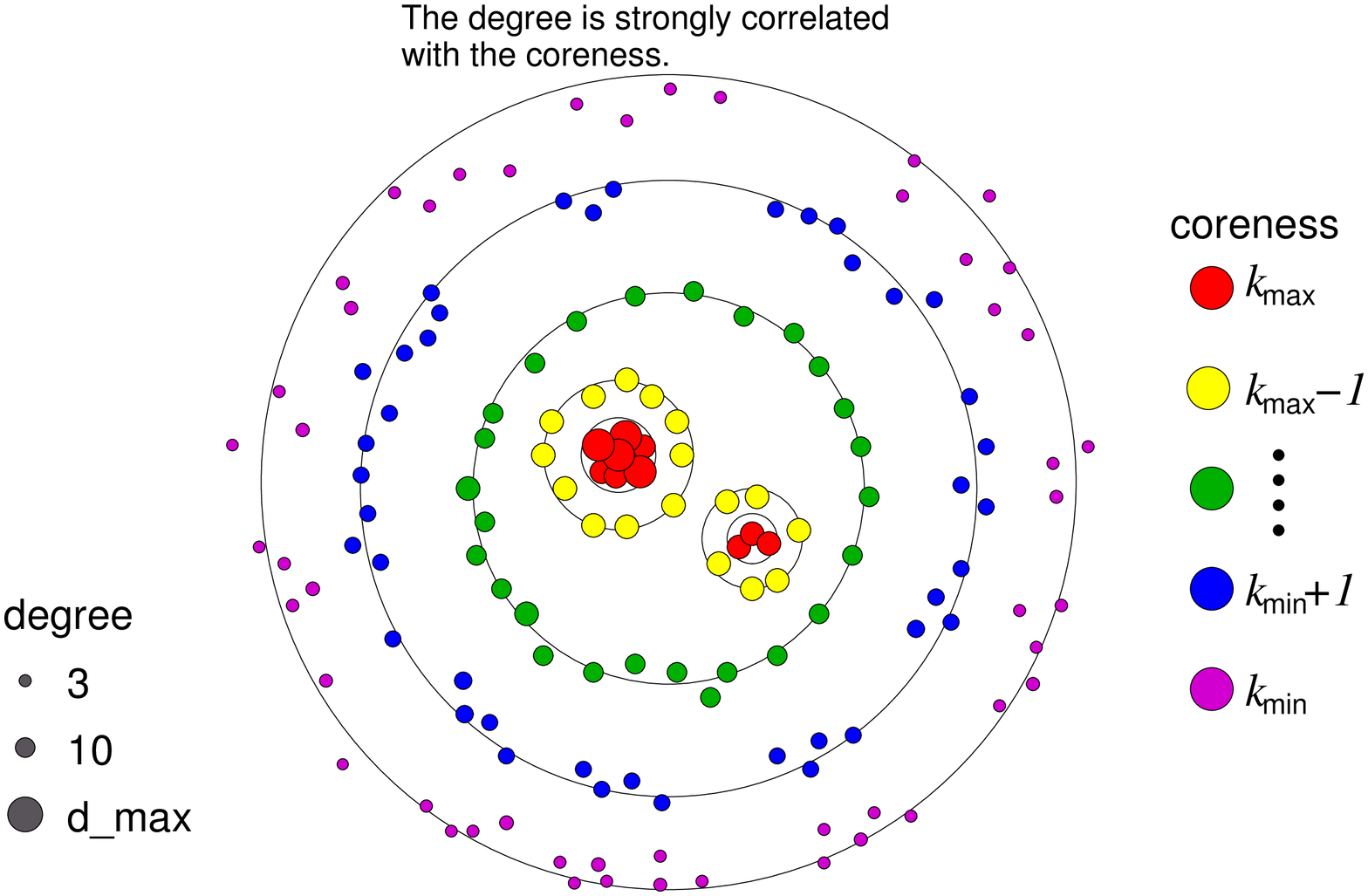}
\includegraphics[angle=-90,width=75mm]{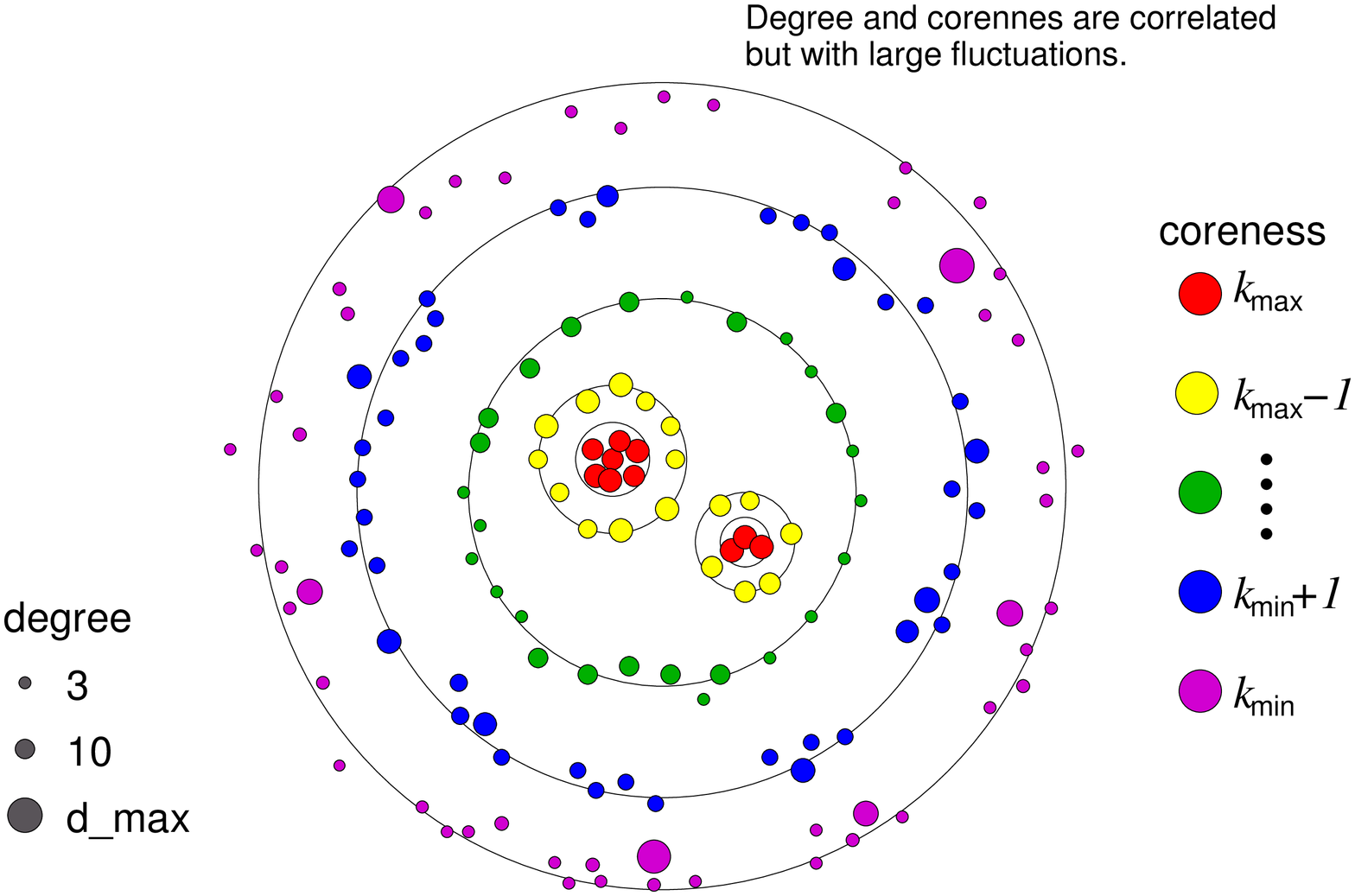} \\
\hline
\end{tabular}
\end{center}
\caption{ The two figures show different correlation properties
between the coreness and the degree of the nodes.  On the left, we
report a graph with strong degree-coreness correlation: the size of
the nodes grows going from the periphery to the center, in strong
correlation with the coreness. The right-hand drawing shows a graph in
which there is the degree-coreness correlations are blurred by large
fluctuations, as stressed by the presence of some hubs in the external
shells.}
\label{k-core5}
\end{figure}

{\bf 2)}
\textit{\textbf{Shell Clusters:}} The angular distribution of
  vertices in the shells is not completely homogeneous. Fig.\ref{k-core4}
  shows that clusters of vertices can be observed. The idea is that of
  grouping together all nodes of the same coreness set that are directly
  linked in the original graph and of representing them close one to another
  in the shell. Thus, a shell is divided in many angular sectors, each one
  containing a cluster of vertices. This feature allows to figure out at a
  glance if the coreness sets are composed of a single large connected
  component rather than divided into many small clusters, or even if there
  are isolated vertices (i.e. disconnected from all other nodes
  in the shell, not from the rest of the $k$-core!).

{\bf 3)}
\textit{\textbf{Degree-Coreness Correlation:}} Another
  property that can be studied from the obtained layouts is the correlation
  between the degree of the nodes and the coreness value. In fact, both 
  quantities are centrality measures and the presence or the absence of
  correlations between them is a very important feature characterizing
  a network's topology.  The nodes displayed in the most internal shells are
  those forming the central core of the network; the presence of
  degree-coreness correlations then corresponds to the fact that
  the central nodes are most likely high-degree hubs of the network. 
  This effect is indeed observed in many real communication networks
  with a clear hierarchical structure, as
  the Internet at the Autonomous System level or the World Wide
  Air-transportation network. On the contrary, the presence of hubs in
  external shells is typical of networks without a clear global
  hierarchical structure as the World-Wide Web or the
  Internet Router Level. In this case, emerging star-like
  configurations appear with high degree vertices connected only to
  very low degree vertices. These vertices are rapidly pruned out in
  the k-core decomposition even if they have a very high degree, 
  leading to the presence of local hub in the
  external k-shells, as in Fig.~\ref{k-core5}. 

{\bf 4)}
\textit{\textbf{Edges:}} The visualization shows only a homogeneously
randomly sampled fraction of the edges. We can tune the
percentage of drawn edges in order to get the better trade-off between
the clarity of visualization and the necessity of giving information
on the way the nodes are mainly connected.  Edge-reduction techniques
can be implemented to improve the algorithm's capacity in representing
edges; however, a homogeneous sampling does not alter the extraction
of topological information, ensuring a low computational cost.
Finally, the two halves of each edge are colored with the color of
the corresponding extremities to make more evident the connection
 among vertices in different shells.

{\bf 5)} \textit{\textbf{Disconnected components:}} The fragmentation
of any given k-shell in two or more disconnected components is represented by
the presence of a corresponding number of circular shells with different centers
(Fig.~\ref{k-core2}).  The diameter of these circles is related with
the number of nodes of each component and modulated by the $\gamma$ parameter
(Eq.~\ref{y_vertex_II}). The distance between components is controlled
by the $\delta$ parameter (Eq.~\ref{x_center}).

  
In summary, the proposed algorithm makes possible a direct, visual
investigation of a series of properties: hierarchical structures of
networks, connectivity and clustering properties inside a given
shell; relations and interconnectivity between different levels of
the hierarchy, correlations between degree and coreness,
i.e. between different measures of centrality.

\section{Results from computer-generated and real networks}\label{results} 

In the following we provide specific examples in which the use of the
proposed visualization algorithm readily allows the identification of
characteristic fingerprints and hierarchies in a set of real and
computer generated networks. In particular, the visualization allows
to identify the lack of hierarchy and structure of the
basic Erd\"os-R\'enyi random graph. Similarly the time correlations
present in the Barab\'asi-Albert network find a clear
fingerprint in our visualization layout. A further
interesting example is the identification of the different
hierarchical arrangement of the Internet network when visualized at
the Autonomous system level and the Router level. These examples
provide an illustration of the use and capabilities of the proposed
algorithm in the analysis of large sparse graphs. The 
parameters 
are set to the values $\epsilon=0.18$, $\delta=1.3$ and
$\gamma=1.5$, which provide a readable layout at the definition allowed in the
present paper format.

\begin{figure}[t] 
\begin{center} 
\begin{tabular}{cc}
  \parbox{82mm}{
  \parbox[b]{44mm}{\includegraphics[height=36.5mm,angle=0]{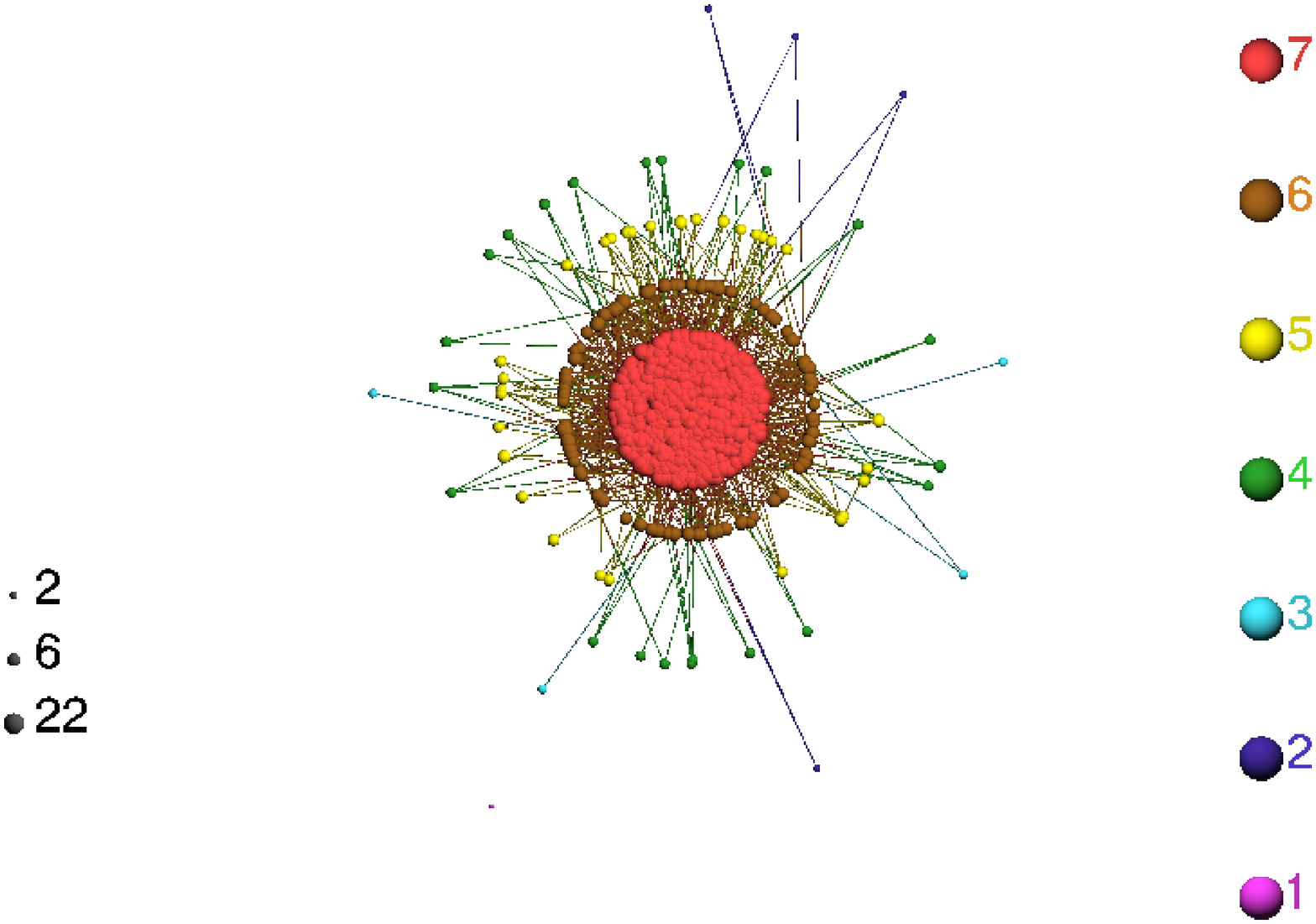} }
  \hskip 1.1mm \parbox[b]{30mm}{\includegraphics[height=36.5mm,angle=0]{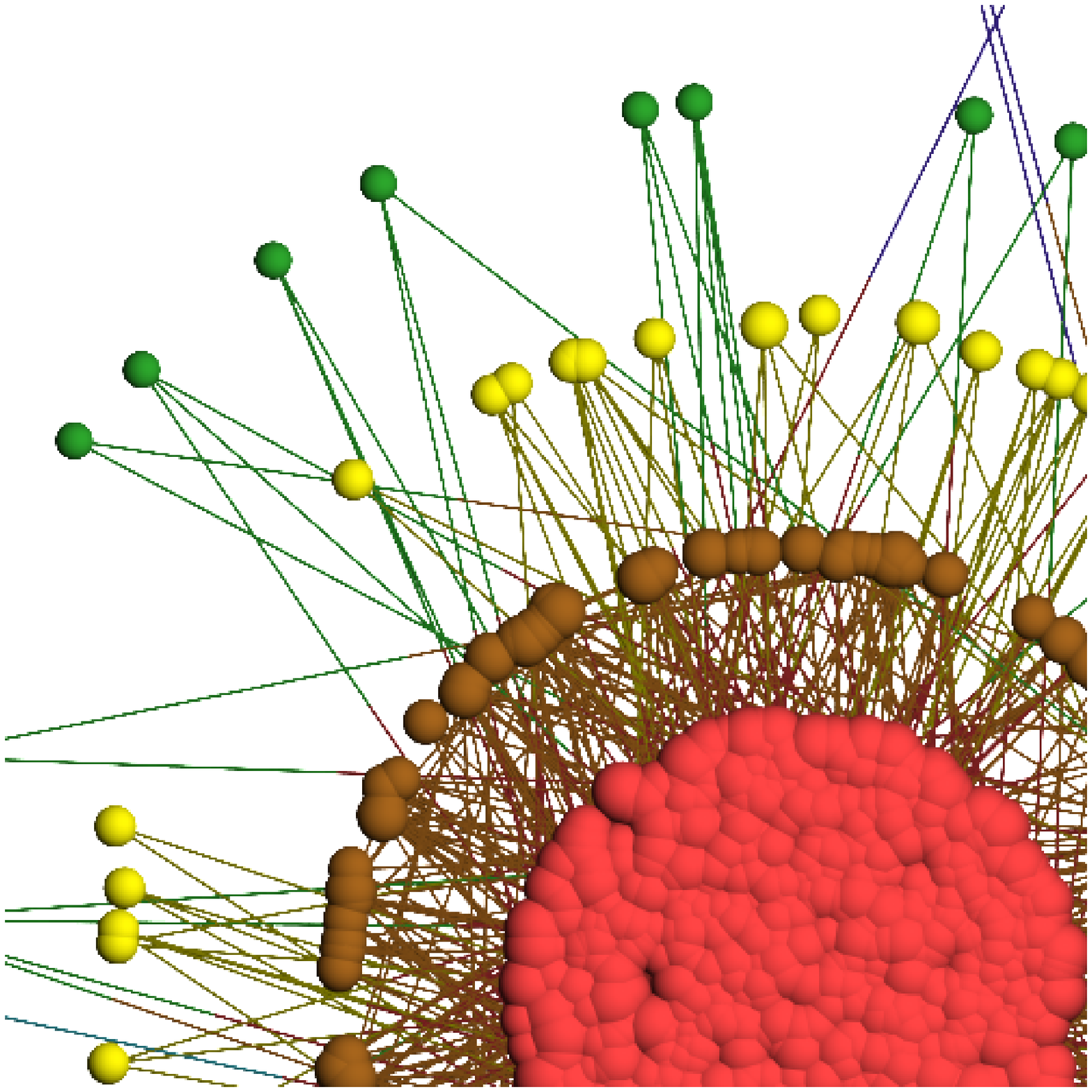} } } &
  \parbox{82mm}{
  \parbox[b]{44mm}{\includegraphics[height=36.5mm,angle=0]{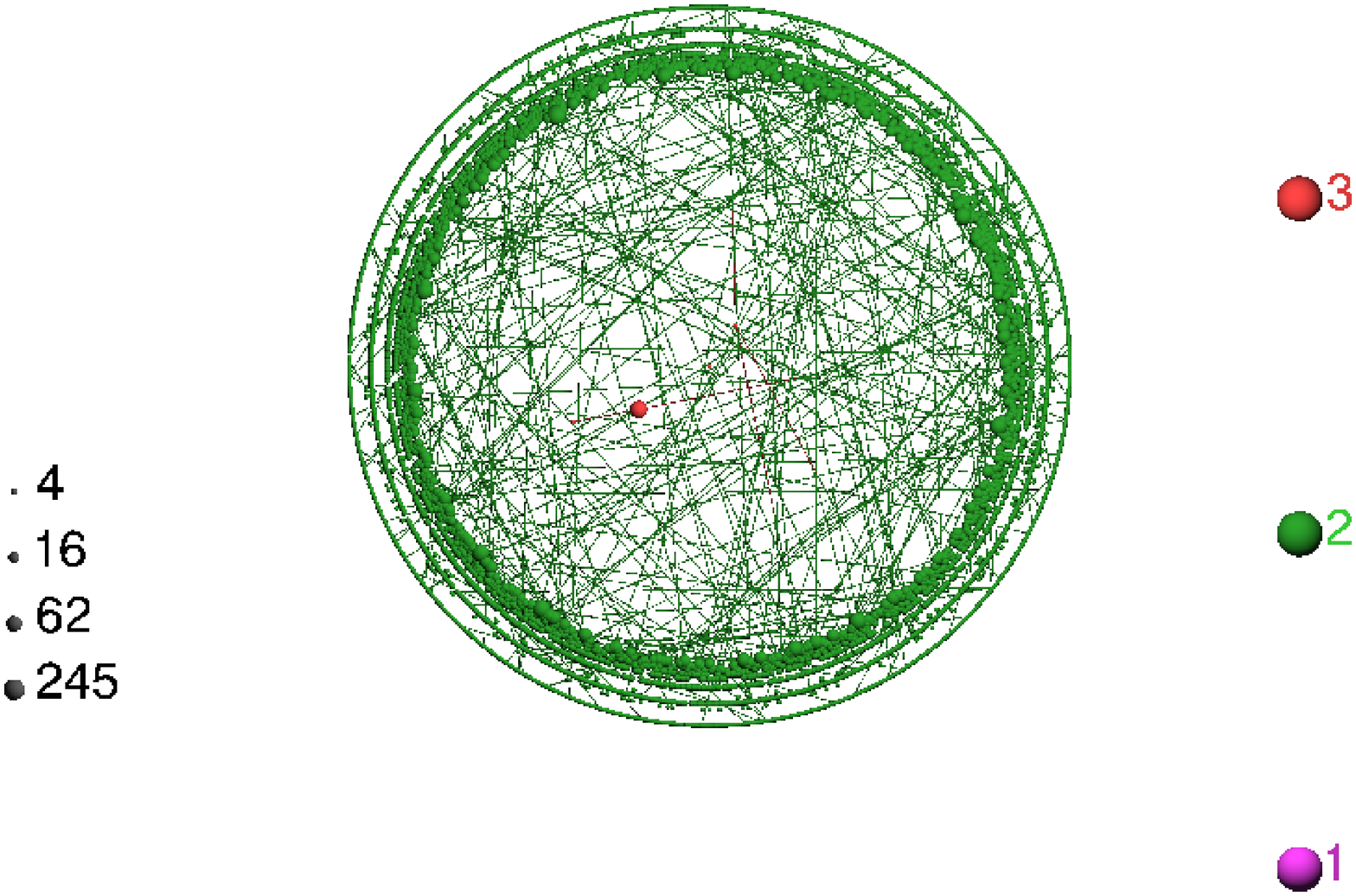} }
  \hskip 1.1mm \parbox[b]{30mm}{\includegraphics[height=36.5mm,angle=0]{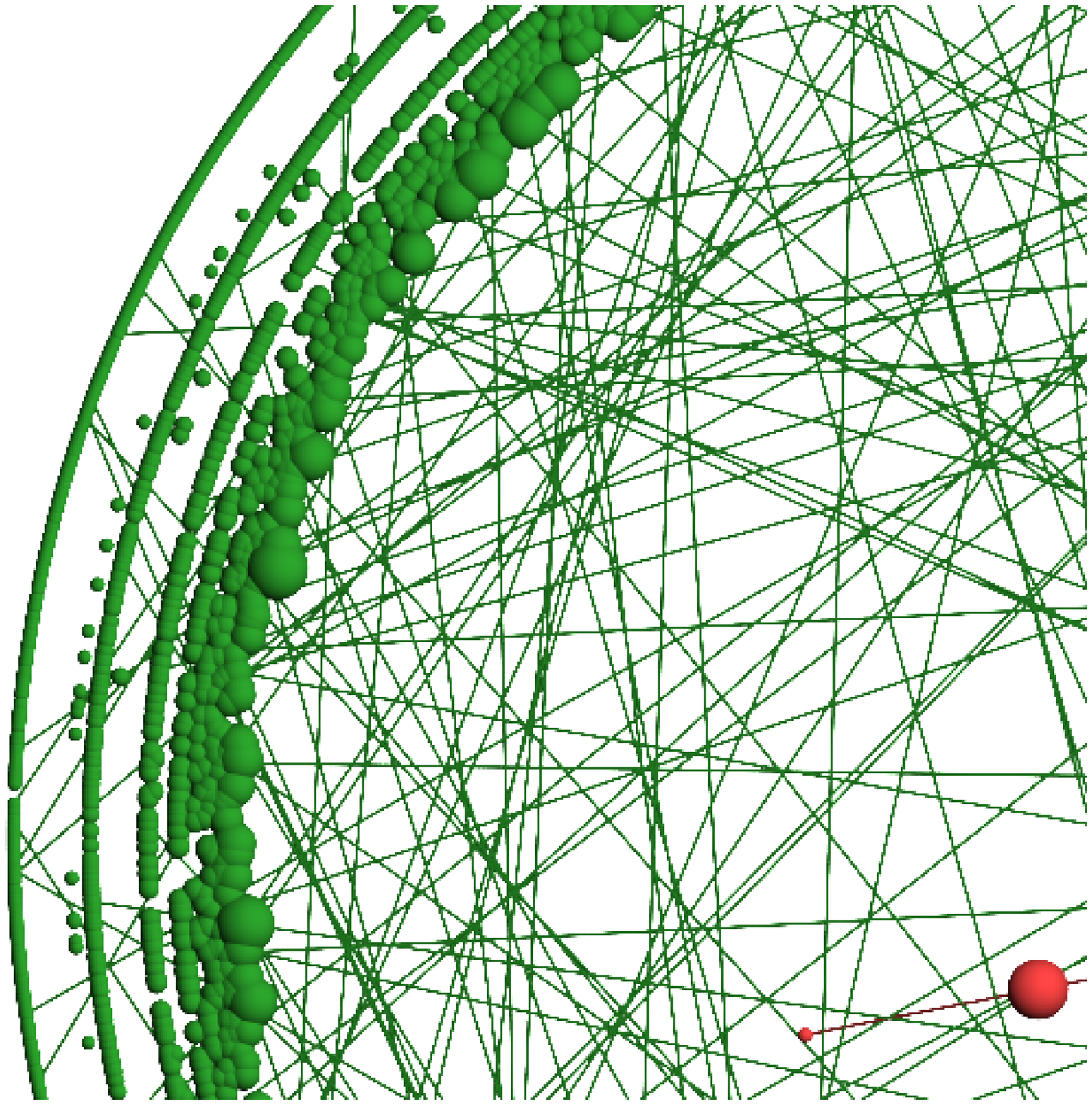} } } \\
  \includegraphics[width=82mm,angle=0]{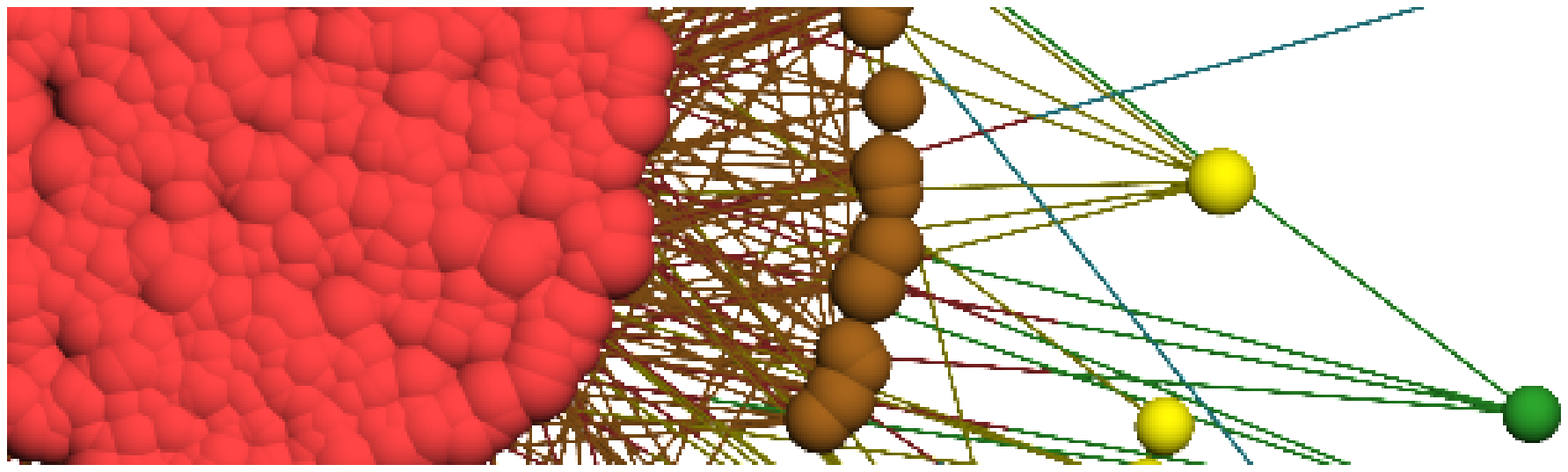} &
  \includegraphics[width=82mm,angle=0]{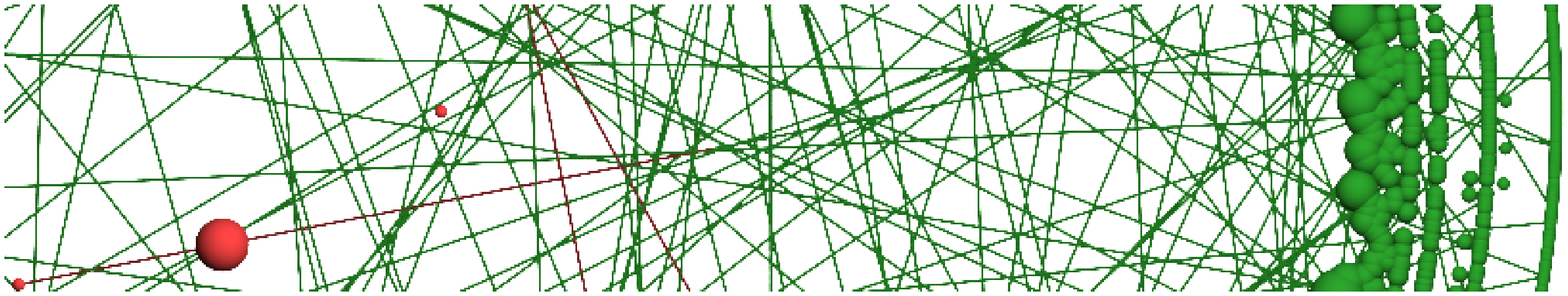} \\ 
\end{tabular}
\end{center} 
\caption{
Graphical representation of an ER network
with average degree $10$  and 1000 vertices (left) and of a 
BA network with $m=2$ and $n=10000$ (right).}
\label{k-core_graph_BA_10000} 
\label{k-core_graph_ER_1000} 
\end{figure} 

\subsection{Visualization and analysis of computer-generated graphs} 
In this section we want to provide the visualization of a set of
computer generated networks generally used in the literature to 
model large scale graphs. We will show that the proposed 
algorithm provides a very intuitive visualization of the
difference between the models and the real networks. In this
perspective, the $k$-core decomposition appears as a suitable tool in
the examination and validation of network models.

 
The Erd\"os-R\'enyi (ER) model \cite{ErdosRenyi59}, with poissonian
degree distribution,  is a typical
example of graphs with a characteristic value for the degree (the
average value $\langle d \rangle$).  Since an ER graph can consist of
more than one connected component, we consider only the largest of
these components. An instance of the visualization of Erd\"os R\'enyi
random graphs is provided by Figure~\ref{k-core_graph_ER_1000}: the
maximum coreness is clearly related to the average degree
$\langle d\rangle = 10$.  The large central mass is the result of the
very homogeneous topology; the vertex degrees have only small
fluctuations, thus most vertices  
belong to the same $k$-core that is also the highest. 
   
Since many real-world networks have been shown to display a very heterogeneous
topology as measured by broad degree distributions, many models and mechanisms
have been proposed to construct heterogeneous networks. The most famous is the
Barab\'asi-Albert (BA) model~\cite{sf99}, which considers growing networks
according to the preferential attachment mechanism: each new vertex is
connected to $m$ already existing vertices chosen with a probability
proportional to their starting degree.  
This model produces graphs with
power-law degree distributions, thus characterized by a very large variety of
degree values.
Such a graph, with $m=2$, shown in
Fig.~\ref{k-core_graph_BA_10000} produces a quite peculiar
decomposition.  Indeed, although this graph displays a very
heterogeneous vertex degree distribution, its $k$-core decomposition is
trivial; only few layers at very small coreness are visible. The
construction mechanism provides a simple explanation. Each new vertex
enters the system with degree $m$, but at the following time steps new
vertices may connect to it, increasing its degree. Inverting the
procedure, we obtain exactly the $k$-core decomposition. The minimum
degree is $m$, therefore all coreness sets $C_c$ with $c < m$ are
empty. Recursively pruning all vertices of degree $m$, one first removes
the last vertex, then the one added at the preceding step, whose degree
is now reduced to the initial value $m$, and so on, up to the initial
vertices which may have larger degree. Hence, all vertices except the
initial ones belong to the coreness set of coreness $m$.  
This somewhat pathological property holds for all growing networks with
fixed initial number of links for new vertices. 
Simple variations of the basic algorithm and the introduction of 
stochasticity in the growing procedure result in more complicate structures.
  


\subsection{Visualization of real networks} 

\begin{figure}[t] 
\begin{center} 
\hspace{-3mm}\noindent \begin{tabular}{c}
  \parbox{84mm}{
  \parbox[b]{45mm}{\includegraphics[height=36.5mm,angle=0]{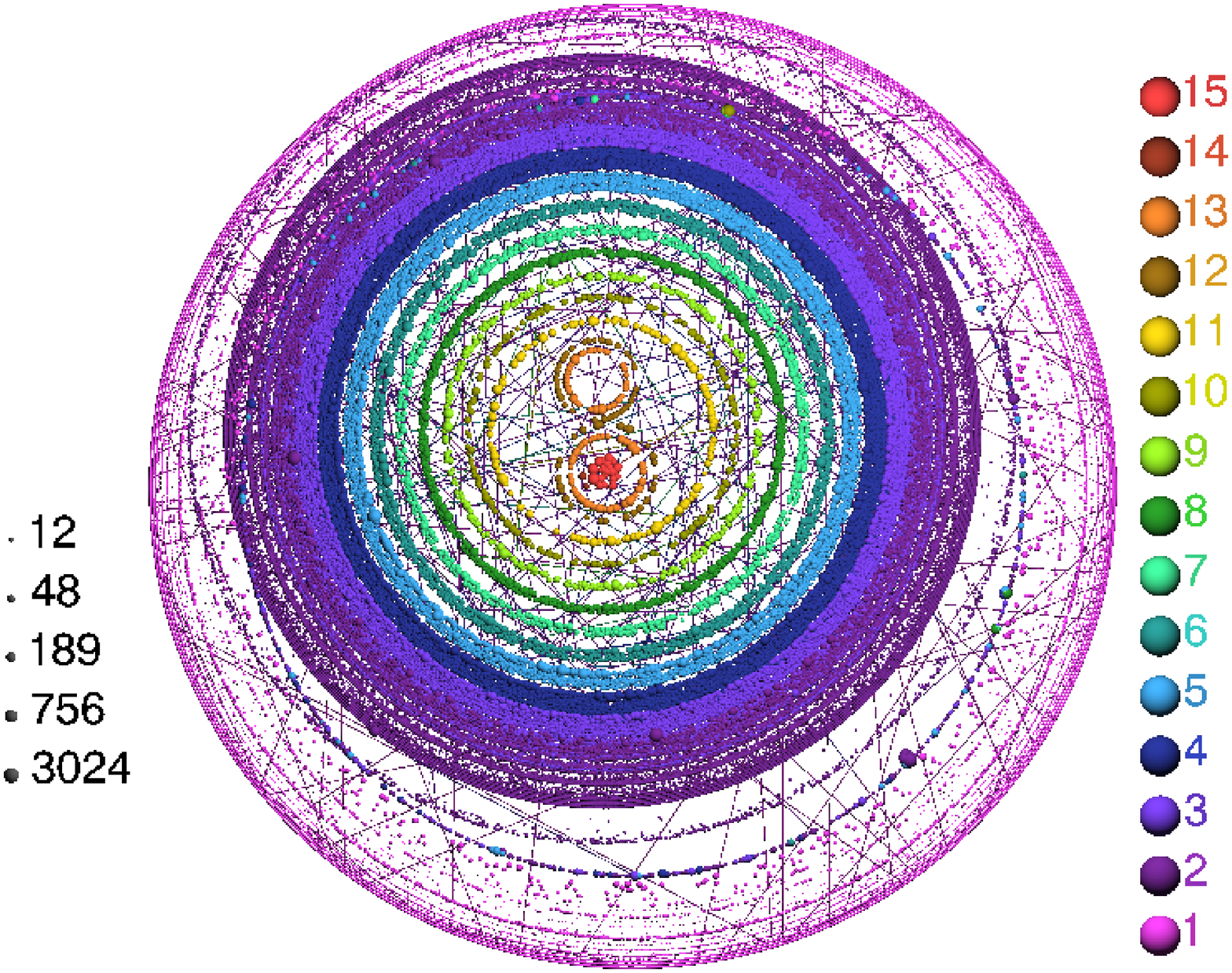} }
  \hskip 1.1mm \parbox[b]{30mm}{\includegraphics[height=36.5mm,angle=0]{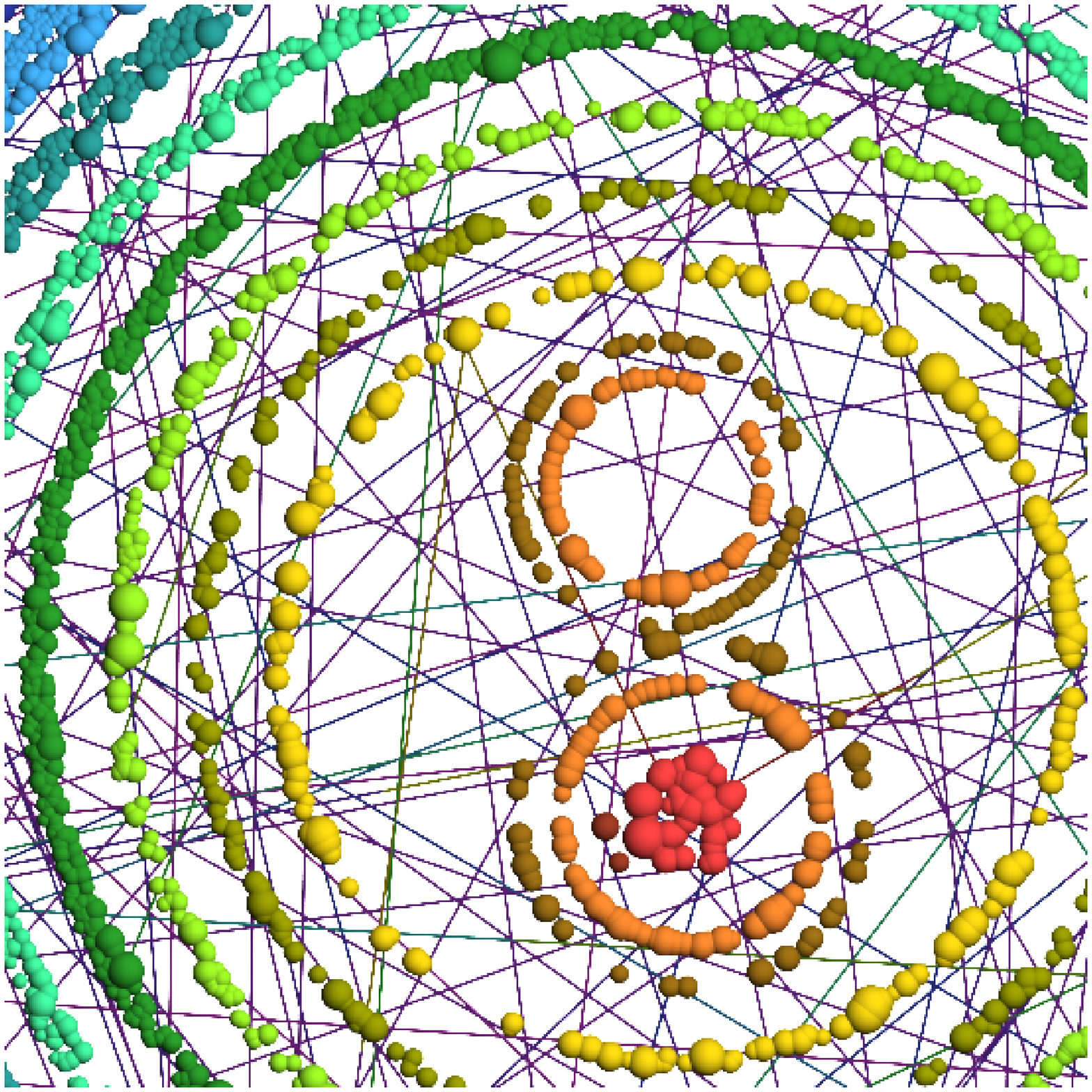} }
  } 
 \\
  \includegraphics[width=82mm,angle=0]{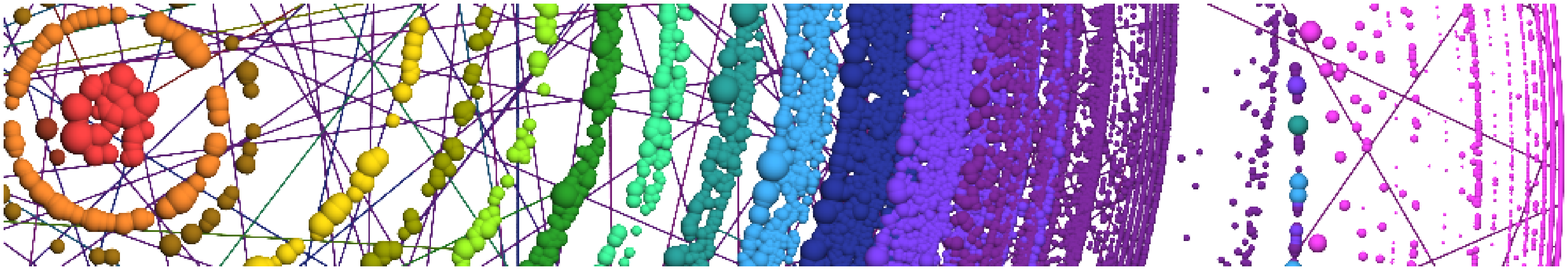} 
\end{tabular}
\end{center} 
\caption{
Graphical representation of a fraction of the {\tt .fr} domain of Web.
}
\label{k-core_graph_WWW}
\end{figure}

We first present a visualization of a portion of the {\tt .fr} domain of 
the World Wide Web (WWW). Its graph is composed by one million pages.
This network, whose visualization is presented in
Figure~\ref{k-core_graph_WWW},
is particularly interesting because at the $12$-core level two disconnected 
components emerge. Note that, 
since the actual definition of $k$-cores concerns undirected graphs,
we consider here the WWW as undirected.
%

We also provide the visualization of networks representing Internet at
various granularity levels. More precisely, we 
consider graphs of Internet at the autonomous system  and  
router level. The autonomous system level is represented by collected 
routes of {\em Oregon route-views}~\cite{oregon} project, called AS, 
and its extended version AS+ presented in Chen {\em et 
al.}~\cite{CCGJSW02}, both from May 26, 2001. For the router level, 
we use the graph obtained by an exploration of Govindan and 
Tangmunarunkit~\cite{govindan00heuristics} in 2000, called here IR 
graph, and the IR\_CAIDA graph obtained from the CAIDA 
project~\cite{IR_CAIDA} between April 21st and May 8th, 2003. Both
networks are composed by approximately $200000$ nodes.
The two ASes maps (close to $11500$ nodes each) 
differ mainly in the number of links: the AS+ maps 
were constructed by using informations from peering relationship of 
autonomous systems, obtained from {\em Looking Glass} tools.  These 
tools are maintained by ISPs to troubleshoot routing problems.

\begin{figure}[t] 
\hspace{-3mm}\noindent \begin{tabular}{cc}
  \parbox{84mm}{
  \parbox[b]{45mm}{\includegraphics[height=36.5mm,angle=0]{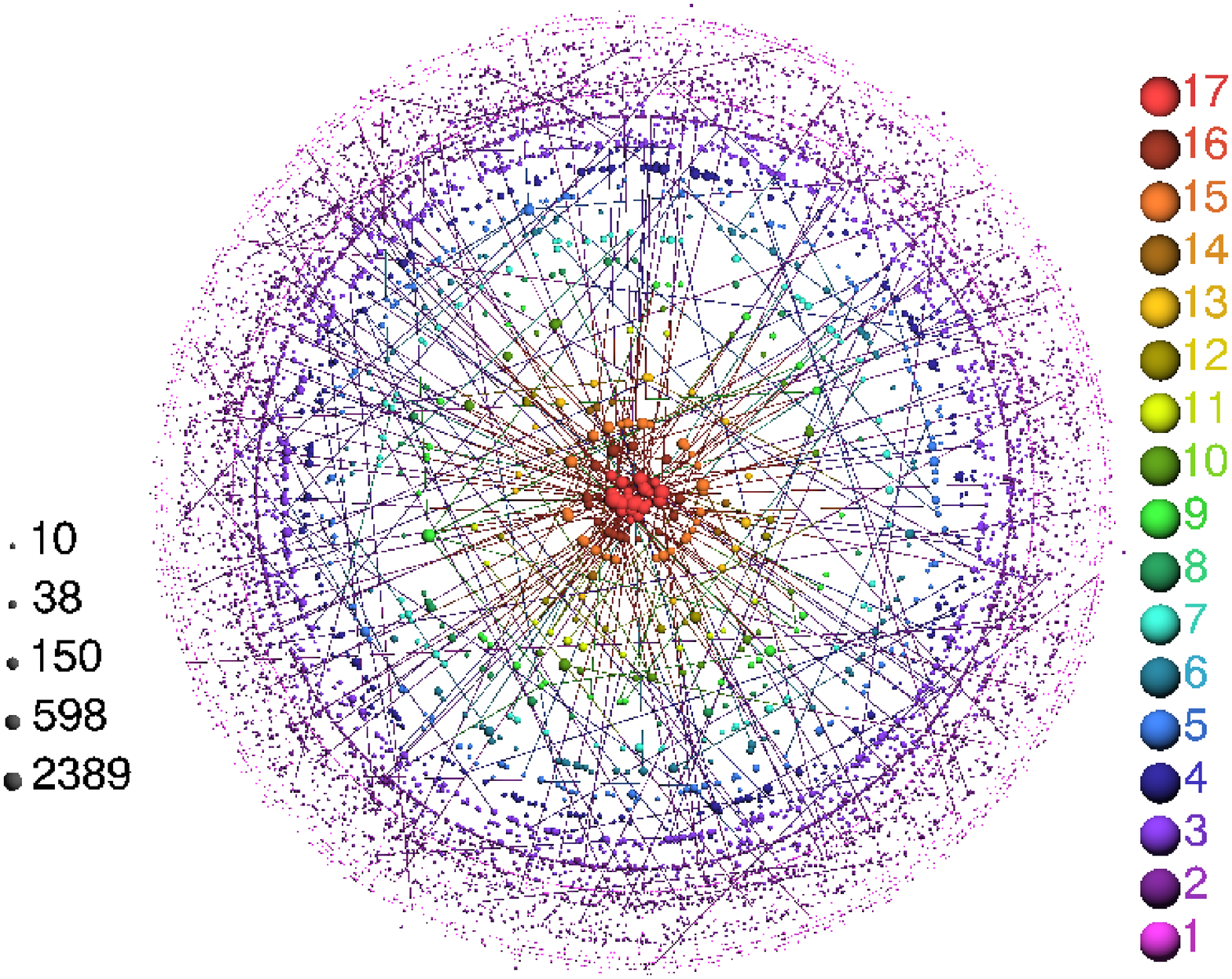} }
  \hskip 1.1mm \parbox[b]{30mm}{\includegraphics[height=36.5mm,angle=0]{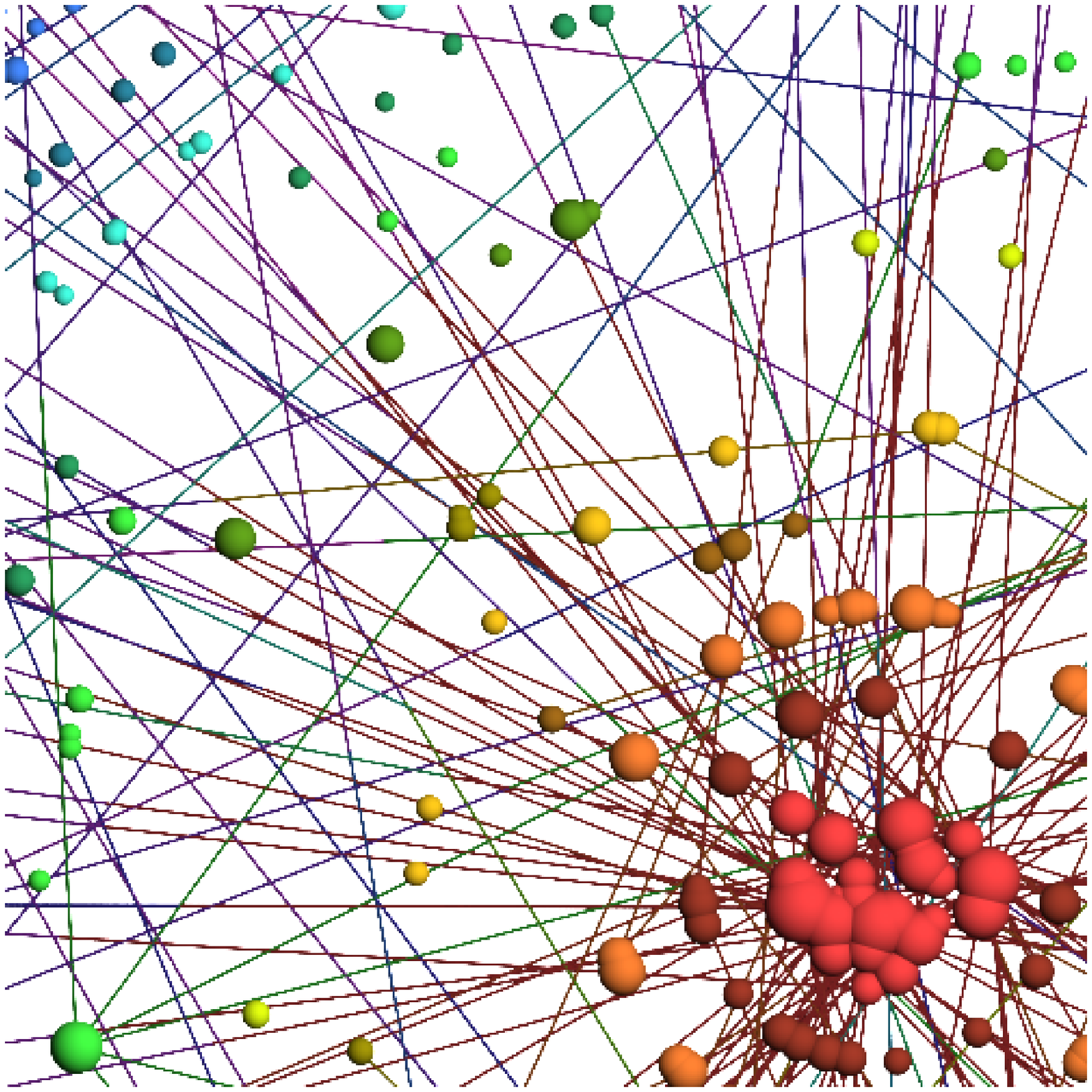} }
  } &
  \parbox{84mm}{
  \parbox[b]{45mm}{\includegraphics[height=36.5mm,angle=0]{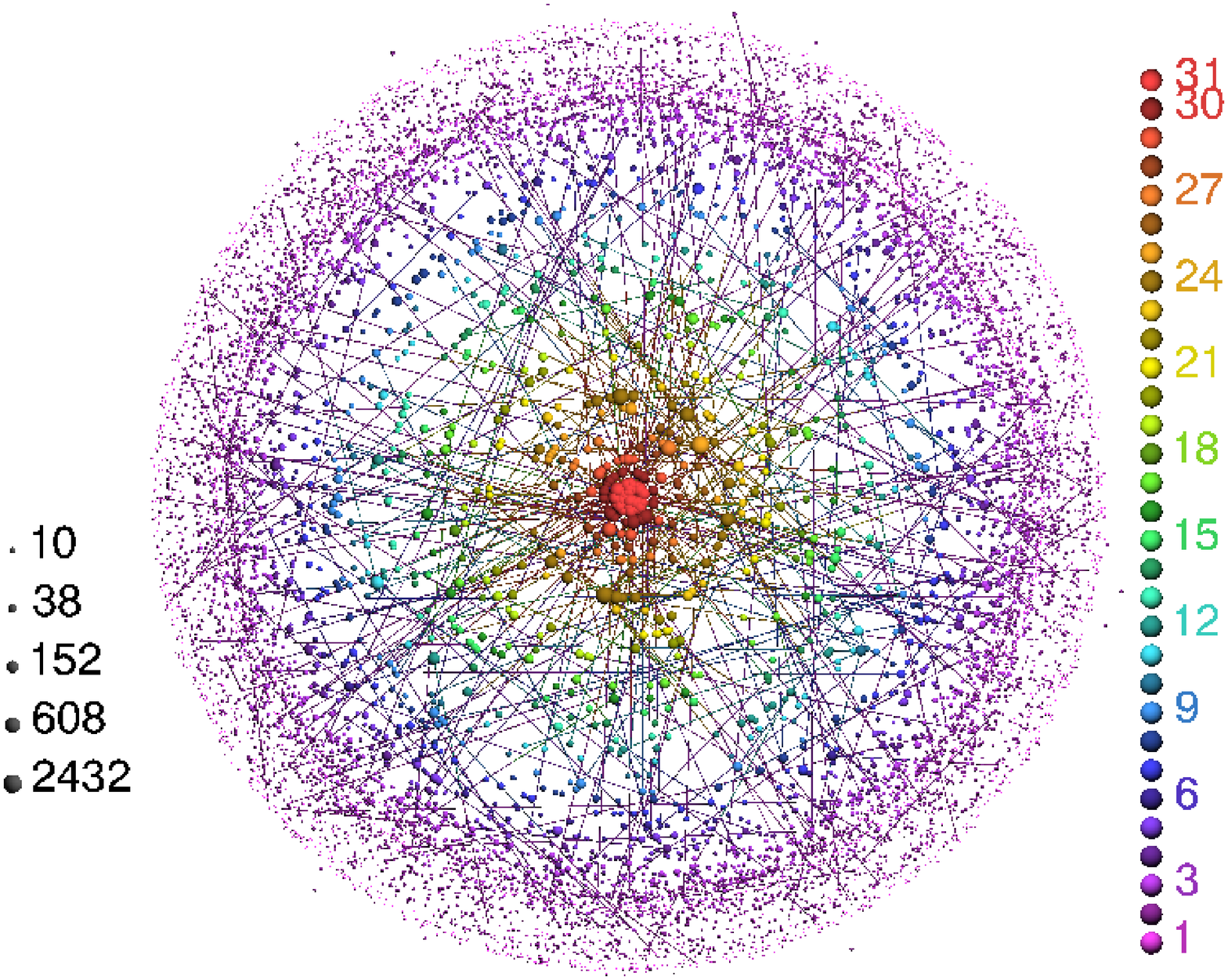} }
  \hskip 1.1mm \parbox[b]{30mm}{\includegraphics[height=36.5mm,angle=0]{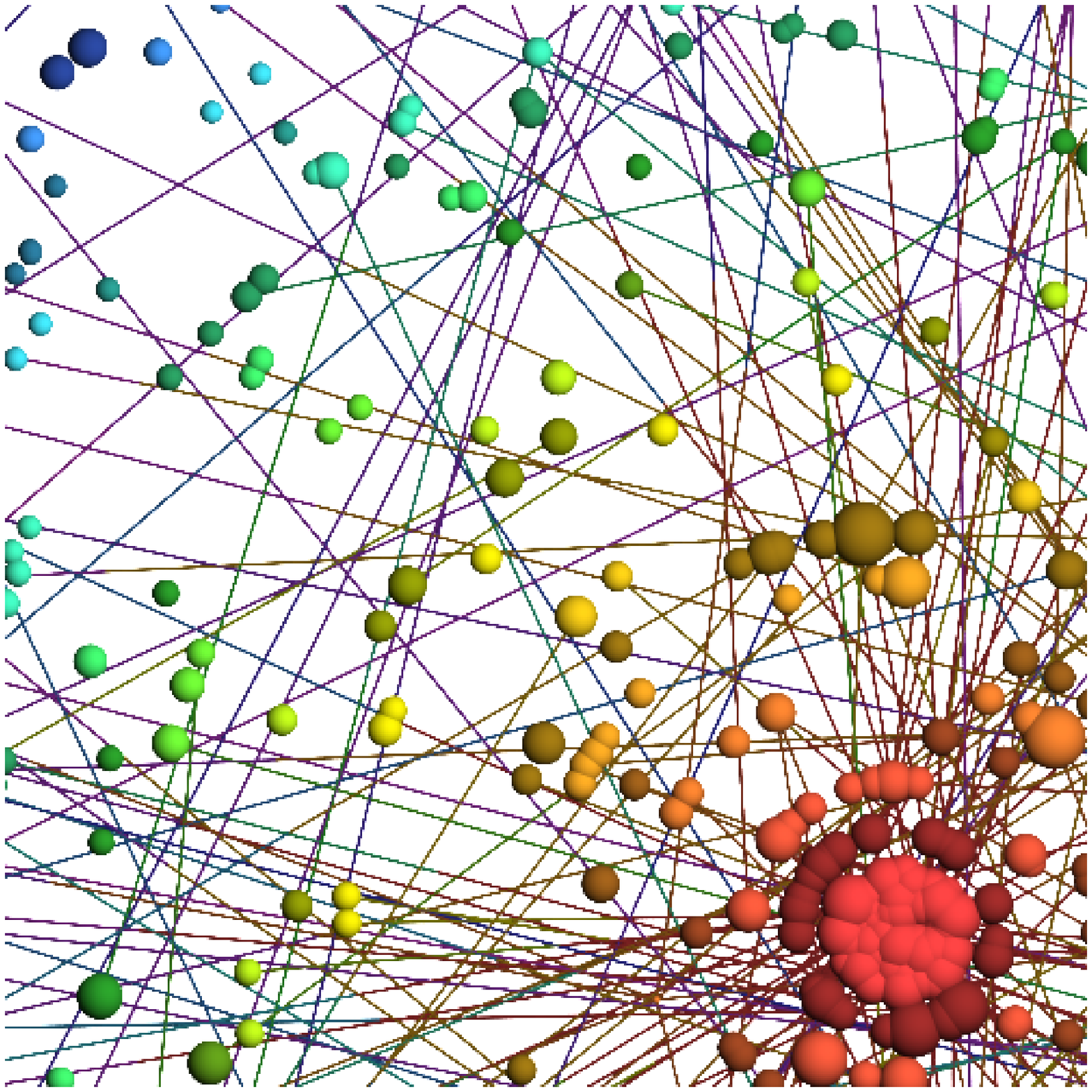} }
  } \\
  \includegraphics[width=84mm,angle=0]{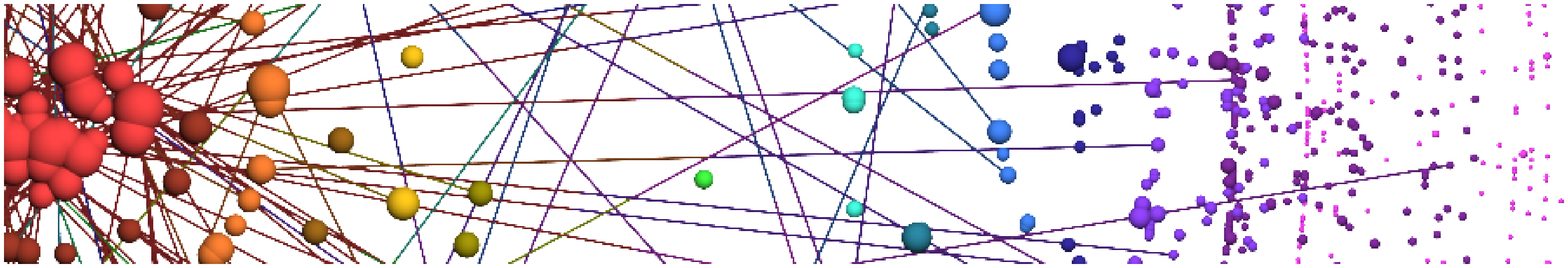}  &
  \includegraphics[width=84mm,angle=0]{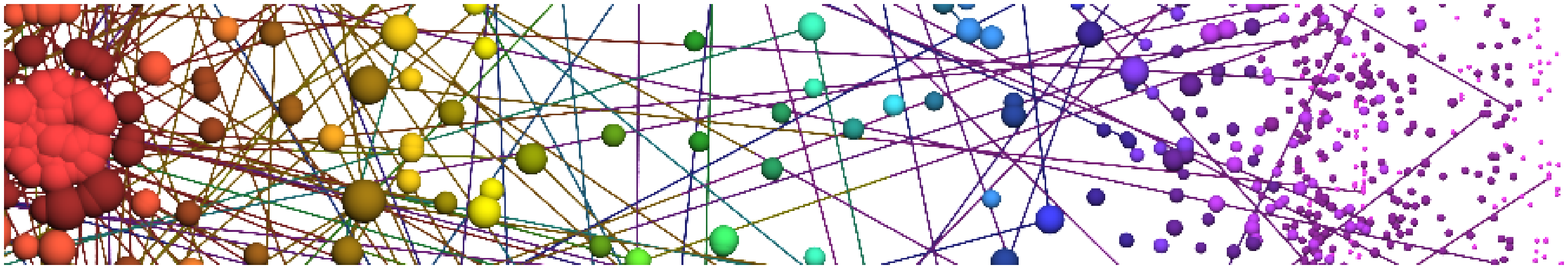} \\
\end{tabular}
\caption{Graphical representation of the AS (left)
and AS+ (right) graphs.}
\label{k-core_graph_AS} \label{k-core_graph_AS+} 
\end{figure}

Figure~\ref{k-core_graph_AS+} displays the
representation of
two different maps of the autonomous system graphs (AS and AS+).
All coreness layers are populated, and, any given
k-shell, the vertices are distributed on a relatively large range
of the radial coordinate, which means that their neighborhoods are
variously composed.  It is worth noting that the coreness and the
degree are very correlated, with a clear hierarchical structure. Links
go principally from one coreness set to another, although there are of
course also intra-layer links. The hierarchical structure exhibited by
our analysis of the autonomous system level is a striking property;
for instance, one might exploit it for showing that in the Internet
high-degree vertices are naturally (as an implicit result of the
self-organizing growth) placed in the innermost structure.

\begin{figure}[t] 
\hspace{-3mm}\begin{tabular}{cc}
  \parbox{84mm}{
  \parbox[b]{45mm}{\includegraphics[height=36.5mm,angle=0]{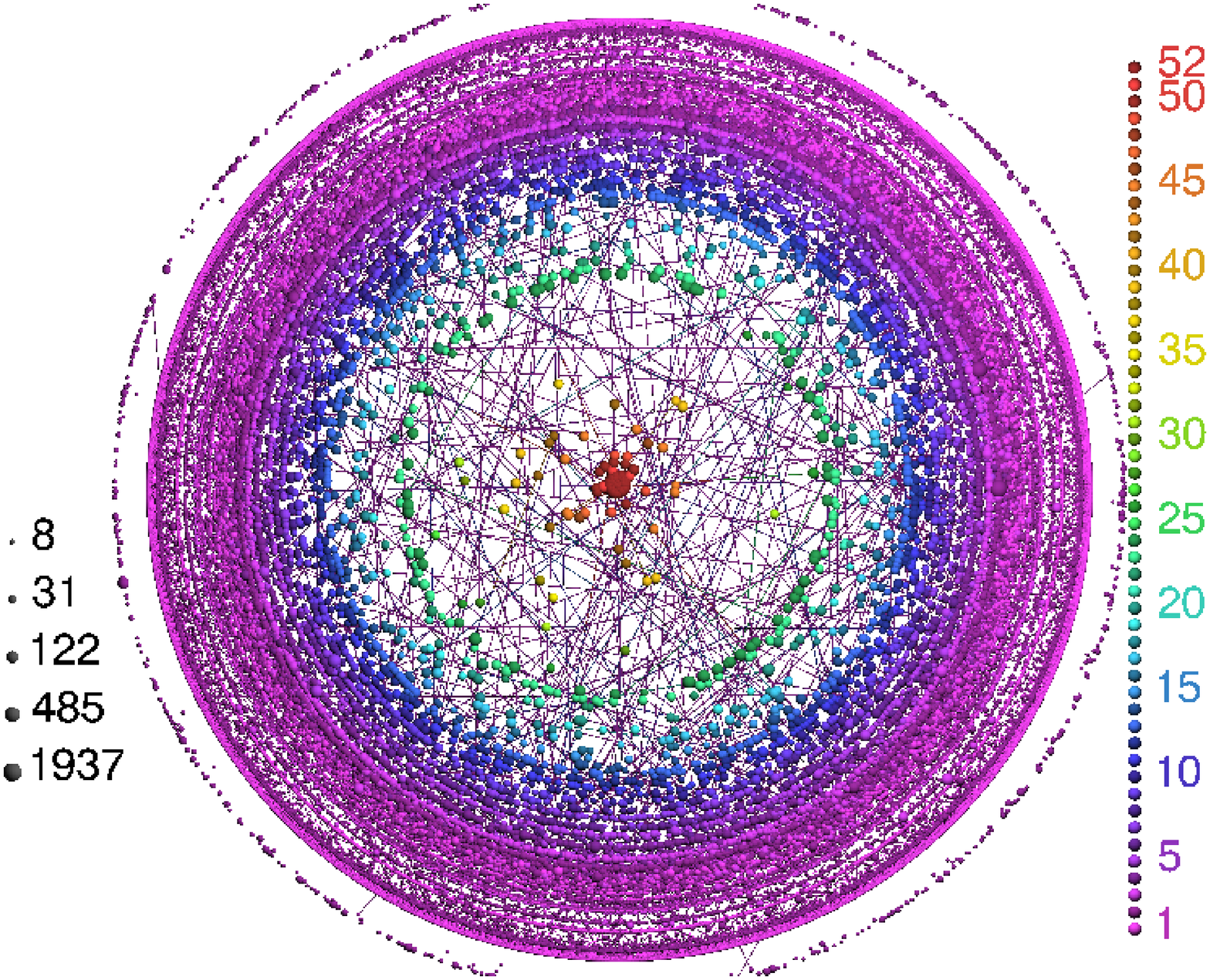} }
  \hskip 1.1mm \parbox[b]{30mm}{\includegraphics[height=36.5mm,angle=0]{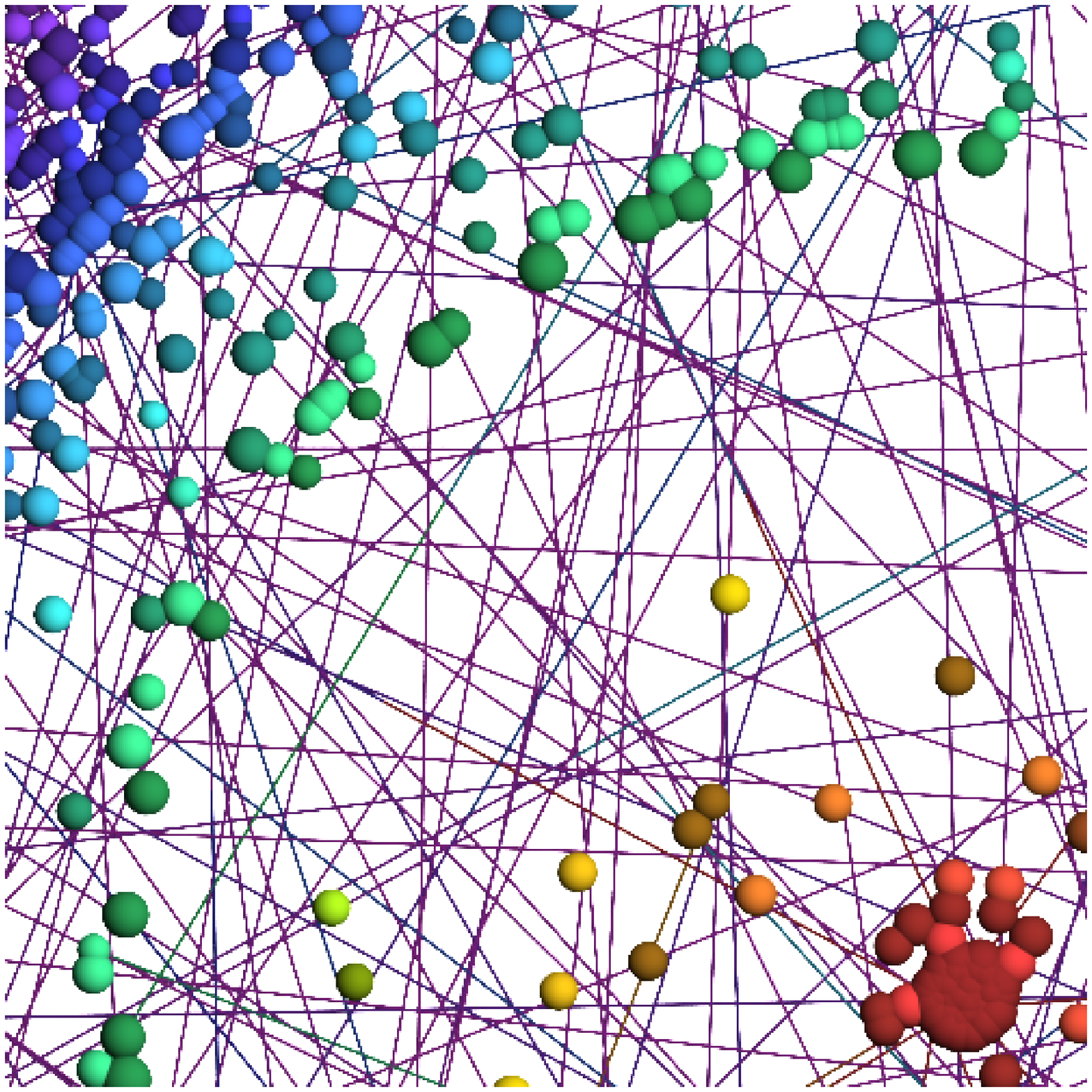} }
  } &
  \parbox{84mm}{
  \parbox[b]{45mm}{\includegraphics[height=36.5mm,angle=0]{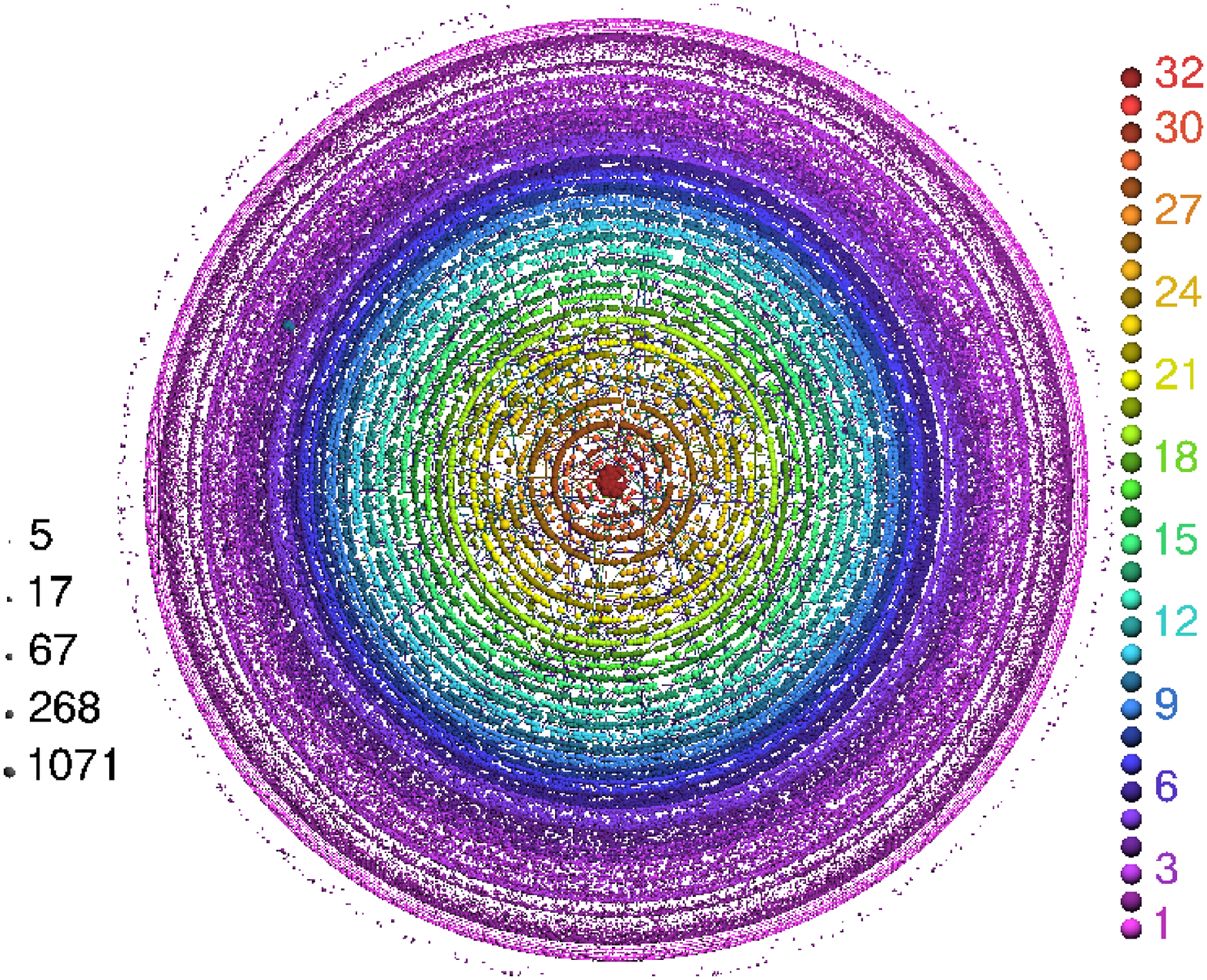} }
  \hskip 1.1mm \parbox[b]{30mm}{\includegraphics[height=36.5mm,angle=0]{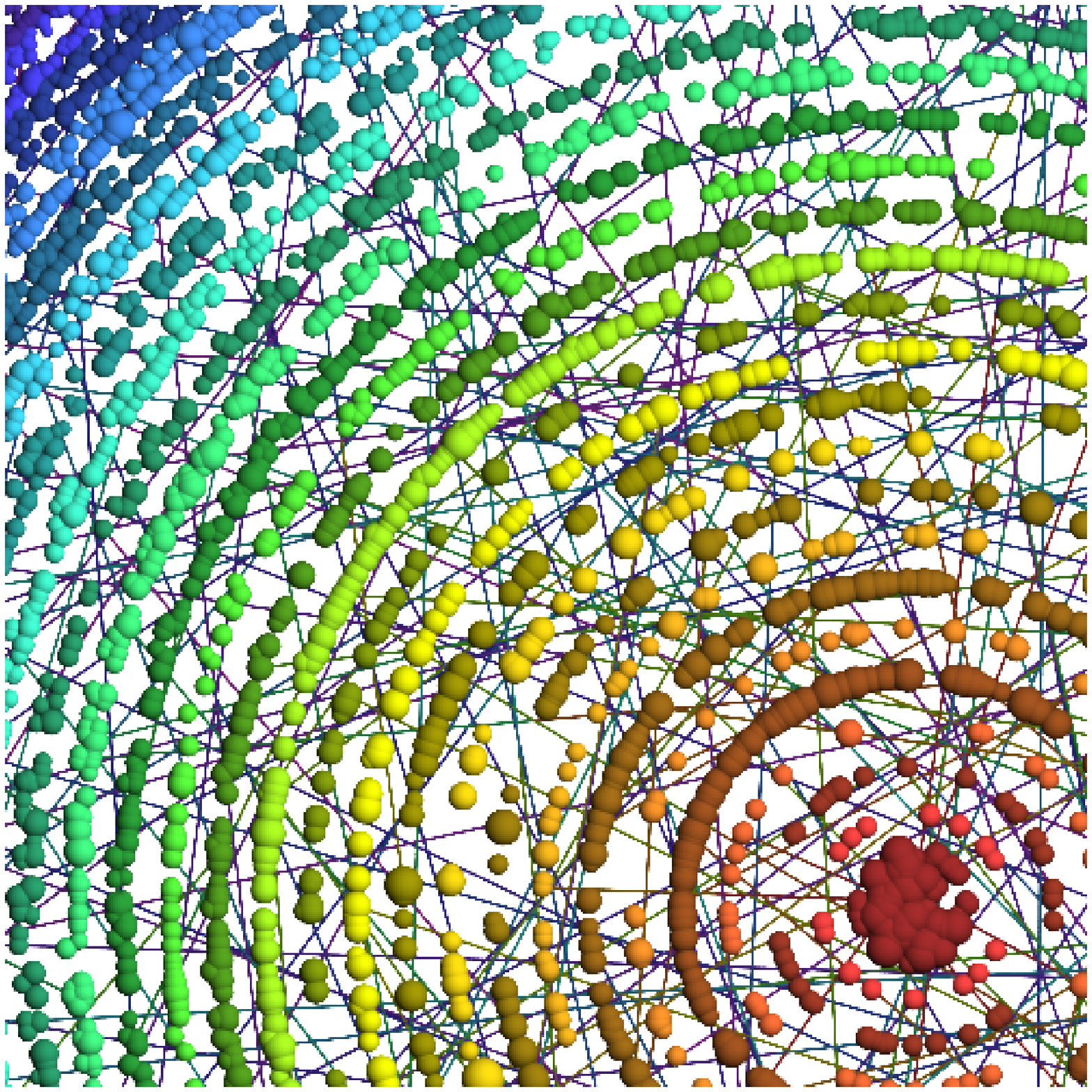} }
  } \\
  \includegraphics[width=84mm,angle=0]{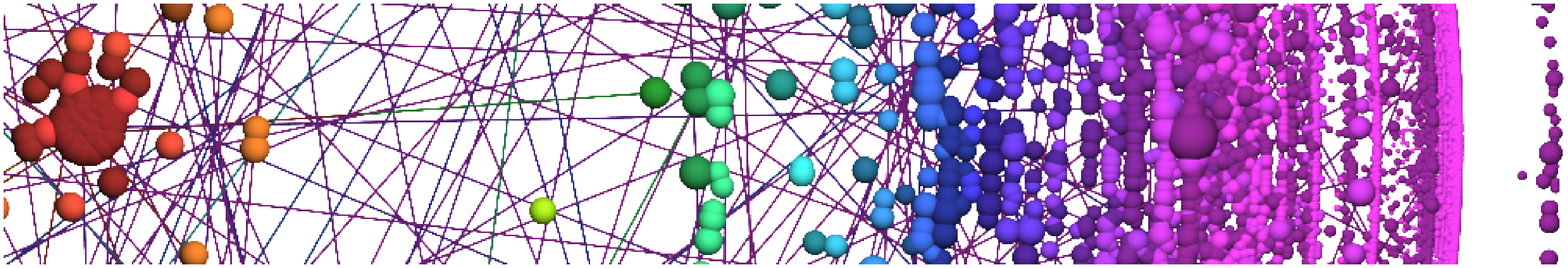} &
  \includegraphics[width=84mm,angle=0]{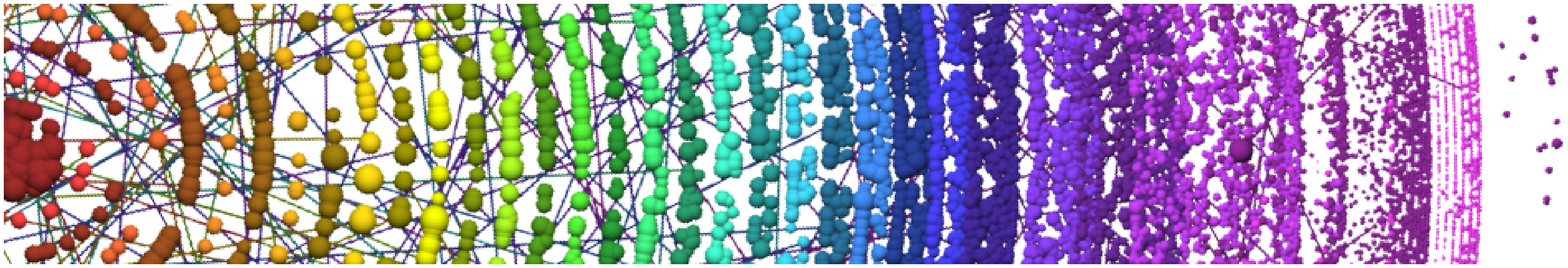} \\
\end{tabular}
\caption{ Left: Graphical representation of the IR (left)
and IR\_CAIDA (right) graphs.}
\label{k-core_graph_IR} \label{k-core_graph_IR_CAIDA} 
\end{figure}

At high resolution, i.e. at the router (IR) level, Internet's
properties are less structured, as shown in
Figure~\ref{k-core_graph_IR_CAIDA}, in which a completely different
scenario emerges: external layers, of lowest coreness, contain vertices
with large degree. For instance, in the IR graph we find 20 vertices
with degree larger than 100 which have coreness smaller
than 6.  
The
correlation between coreness and degree is thus clearly of a very
different nature in the maps of Internet obtained at different
granularities i.e. routers or autonomous systems.

The lowest $k$-shells, containing vertices that are very external, are
displayed as quite broad shells, meaning that the corresponding vertices
have neighbors with coreness covering a large range of values.  The
larger coreness shells are
thin rings, which means that the neighbors of the
vertices in a given layer have similar coreness.
%

It is worth remarking how the present visualization allows the
distinction of networks which appear very similar on the basis of 
the sole statistical properties. 
Indeed, we can notice that the IR map is quite different 
from the IR\_CAIDA map. This difference likely finds its origin in 
the different exploration methods used to gather the two data sets. 
The IR map has been obtained from one source monitor, using 
{\em source routing} to detect lateral connectivity.  
The IR\_CAIDA map, instead, is the merger of data gathered by several
different probing monitors.  On one hand, it is likely 
that the most central cores of the IR network are composed by routers 
with {\em source routing} activated (approximately $8\%$ of the 
total routers~\cite{govindan00heuristics}).
These routers sample destinations unevenly resulting in a less
regular layout. On the other hand, the IR\_CAIDA map appears to have a 
very regular structure likely due to a more symmetric exploration process.
The obtained layout provides at a glance the evidence for pronounced 
differences in the structural ordering in the two maps, suggesting the 
critical examination and comparison of the two experimental strategies.

\section{Conclusions}\label{concl} 
 
In this paper, we have proposed a general visualization tool for large scale
graphs.  Exploiting $k$-core decomposition, and the natural
hierarchical structures emerging from it, our algorithm yields
a layout that possesses the simplicity of a 2D representation with
a considerable amount of information encoded.  One can easily
read basic features of the graph (degree, hierarchical structure,
position of the highest degree vertices, etc.)  as well as more
entangled features, e.g. the relation between a vertex and the
hierarchical position of its neighbors.  Our results show the
possibility of gaining clear insights on the architecture of many real
world and computer-generated networks by a visualization based on the
rationalization of the corresponding graph.  In conclusion, the
present visualization strategy is a useful tool for a 
distinction between networks with different topological properties and
structural arrangement, but it may be also used for determining if a
certain model is in good agreement with real data, providing a further
interesting tool for models validation.  
Finally, we also provide a publicly available tool for visualizing 
networks~\cite{LANET-VI}.

\noindent{\bf Acknowledgments:}
We  gratefully acknowledge
Fabien Mathieu of LIRMM at Montpellier, France, for providing the .fr
portion of the WWW graph. 
This work has been partially funded by the
European Commission - Fet Open project COSIN IST-2001-33555 and
contract 001907 (DELIS).

\bibliography{biblio}
\bibliographystyle{spiebib}

\end{document}